\documentclass[%
 reprint,
 onecolumn,
superscriptaddress,
 amsmath,amssymb,
 aps,
]{revtex4-2}

\usepackage{subfigure}
\usepackage[T1]{fontenc}
\usepackage{hhline}
\usepackage{multirow}
\usepackage{float}
\usepackage{makecell}
\usepackage{graphicx}% Include figure files
\usepackage{dcolumn}% Align table columns on decimal point
\usepackage{bm}% bold math
\usepackage{comment}
\usepackage{makecell}
\usepackage{xcolor}
\usepackage{mathrsfs}  
\newcommand{\black}{\color{black}}

\begin{document}

\preprint{APS/123-QED}

\title{\black Survey of machine-learning wall models for large-eddy simulation \black}% Force line breaks with \\

\author{Aurélien Vadrot}
\email{aurelien.vadrot@mpe.au.dk}
\affiliation{%
 Department of Mechanical and Production Engineering, Aarhus University, 8200 Aarhus N, Denmark 
}%
\author{Xiang I.A. Yang}
\email{xzy48@psu.edu}
\affiliation{
 Department of Mechanical Engineering, Pennsylvania State University, State College, PA, 16802, USA}
\author{Mahdi Abkar}% 
\email{abkar@mpe.au.dk}
\affiliation{%
 Department of Mechanical and Production Engineering, Aarhus University, 8200 Aarhus N, Denmark 
}%

\date{\today}

\begin{abstract}
This survey investigates wall modeling in large-eddy simulations (LES) using data-driven machine-learning (ML) techniques. To this end, we implement three ML wall models in an open-source code and compare their performances with the equilibrium wall model in LES of half-channel flow at eleven friction Reynolds numbers between $180$ and $10^{10}$.
The three models have ``seen'' flows at only a few Reynolds numbers.
We test if these ML wall models can extrapolate to unseen Reynolds numbers.
Among the three models, two are supervised ML models, and one is a reinforcement learning ML model. 
The two supervised ML models are trained against direct numerical simulation (DNS) data, whereas the reinforcement learning ML model is trained in the context of a wall-modeled LES with no access to high-fidelity data.
The two supervised ML models capture the law of the wall at both seen and unseen Reynolds numbers---although one model requires re-training and predicts a smaller von Kármán constant.
The reinforcement learning model captures the law of the wall reasonably well but has errors at both low ($Re_\tau<10^3$) and high Reynolds numbers ($Re_\tau>10^6$).
In addition to documenting the results, we try to ``understand'' why the ML models behave the way they behave.
Analysis shows that the error of the supervised ML models is a result of the network design and the error in the reinforcement learning model arises due to the present choice of the ``states'' and the mismatch between the neutral line and the line separating the action map.
In all, we see promises in data-driven ML wall models.

\end{abstract}

\maketitle

%\linenumbers

\section{\label{sec:intro} Introduction}

Machine learning (ML) has been used in a wide range of domains in recent years, including image recognition, market analysis, weather forecast, and others.
Computational fluid dynamics (CFD) is not exempt.
ML tools were applied in modeling \cite{vollant2017subgrid,singh2017machine,yang2019predictive,champion2019data,li2021data,kasten2022modeling,du2022deep,sun2020surrogate}, computation \cite{cai2022physics, xiang2021neuroevolution,arzani2021uncovering}, control \cite{rabault2019artificial}, and optimization \cite{viquerat2021direct,huang2021bayesian,mekki2021genetic}.
An overview of the ML applications in the field of fluid dynamics can be found in Refs. \cite{duraisamy2019turbulence,brunton2020machine,zehtabiyan2022data}.
 
Consider turbulence modeling, an old field that dates back to at least Prandtl and his mixing length model \cite{prandtl1926uber}.
In the past 100 years or so, many empirical models have been developed.
In the field of Reynolds-averaged Navier Stokes (RANS), there exist the Spalart-Allmaras model \cite{spalart1992one}, the SST $k-\omega$ model \cite{menter1992improved}, the Full Reynolds Stress Model \cite{launder1975progress,speziale1991modelling}, among others.
In the field of large-eddy simulation (LES), there exist the Smagorinksy sub-grid scale (SGS) model \cite{smagorinsky1963general}, the Vreman SGS model \cite{vreman2004eddy}, the equilibrium wall model \cite{schumann1975subgrid}, among others.
These empirical models and the ones in Refs. \cite{chien1982predictions,smith1994near,wilcox2008formulation} have survived many independent comparative studies \cite{menter1994assessment,rumsey2011summary,xu2022direct}, are shown to be Galilean invariant, and preserve the known empiricism like the Kolmogorov's theory of small-scale turbulence \cite{kolmogorov1941local} and the logarithmic law of the wall \cite{marusic2013logarithmic}.
Many of these models are now available in commercial and open-source CFD software like Fluent, STARCCM+,  and OpenFOAM, and can be picked up and used, as they are, by anyone.
However, empirical models are not always sufficiently accurate.
Slotnick \emph{et al.} \cite{slotnick2014cfd} noted that the available empirical models fall short in their predictions of separated flows, high-speed flows, and flows with strong heat transfer.

The inadequacies of empirical models motivated the development of ML models.
In this paper, the term ``empirical models'' refers to conventional, white-box turbulence models with analytical forms, and the term ``ML models'' refers to the more recent, black-box turbulence models that usually have no analytical form.
Note that the wording, i.e., ``physics-based models''  and ``data-based models'' in reference to ``empirical models'' and ``ML models'', is not precise: physics like Galilean invariance is a building block of ML models, and empirical models also invoke data (to determine model coefficients, to pick one term over another).
ML models give accurate results for flows in the training dataset, and many have reported successes \cite{ling2016reynolds,wu2018physics,holland2019towards}.
However, these successes have not benefited front-line CFD practitioners as much as they should/could, largely because of the black-box nature of ML models and the resulting difficulty in implementing them in a CFD code.
As a result, conducting comparative studies is very hard.
Comparative studies are important to turbulence models.
A model must survive many comparative studies before it can be picked up and used for predictive modeling.
However, there are few comparative studies for ML models.
The only study in the present literature seems to be the one by Rumsey \emph{et al.} \cite{rumsey2022search}, where they found that the improvements offered by ML models in some flows are often at the expense of degrading behaviors in other flows.
Rumsey {\it et al.} do not favor trading off generality for accuracy.
The trade-off between generality and accuracy is, in principle, one's choice.
However, CFD and fluid dynamics is a field where getting training data is costly \cite{yang2021grid, li2022grid,yeung2022simulation}, and fluid engineers need to handle unseen flows routinely.
These circumstances make trading off generality for accuracy undesirable.
\black
ML is often criticized for its lack of generality and interpretability.
While the method can automatically discover the ideal model from data without the need for prior hypotheses, it often comes at the expense of physical considerations in modeling \cite{legaard2021constructing}.
The "black-box" nature of machine learning makes it difficult to comprehend the relationship between the input and the model's predictions, which limits the ability to generate simple explanations or hypotheses about the data relationships.

%Consequently, the study by Rumsey \emph{et al.} \cite{rumsey2022search} has engendered many doubts about ML models.
\black
Given this limitation, we pursue a comparative analysis of various machine-learning wall models (MLWMs).
Compared to ML RANS models, ML wall modeling is a less explored territory, making a comparative study a lot less daunting.
Before we proceed further, we review the basics of LES, wall modeling, and the recent progress in ML wall models.

LES is a scale-resolve tool.
An LES resolves the large-scale, more energetic motions and models the small-scale, less energetic motions \cite{Pope2000}.
The tool is more cost-effective than direct numerical simulation (DNS) and more accurate than RANS, and is seeing many applications in academic research \cite{goc2021large,abkar2015influence,abkar2017large,stoll2020large,cho2021wall,zhou2020large,chen2012reynolds,yang2016large,yang2016mean} thanks to the ever more powerful HPC systems.
This work concerns LES of boundary-layer flows.
The energetic motions in a boundary layer scale as their distances from the wall.
As a result, to resolve the energetic motions in a boundary layer, the LES grid must scale as $\nu/u_\tau$ in the inner layer and $\delta$ in the outer layer, respectively.
Here, $\nu$ is the kinematic viscosity, $u_\tau$ is the friction velocity, and $\delta$ is the boundary layer height. 
This inner-layer grid resolution requirement is very restrictive at high, practically relevant Reynolds numbers. 
Yang \emph{et al.} \cite{yang2021grid} estimated the grid-point requirements for wall-resolved LES (WRLES) and wall-modeled LES (WMLES) of a flat-plate boundary layer, where the inner layer is resolved and modeled respectively.
According to them, the grid-point requirements for WRLES and WMLES are $N\sim Re_{L_x}^{1.86}$ and $N\sim Re_{L_x}^{1.0}$, where $Re_{L_x}=L_xU_0/\nu$, $L_x$ is the length of the flat plate, and $U_0$ is the freestream velocity.
It follows that near-wall turbulence modeling, or wall modeling, is a necessity for flows at practically relevant Reynolds numbers \cite{chapman1979computational,spalart1997comments,choi2012grid}.

\black
There are two strategies for WMLES: hybrid RANS/LES- and wall-shear stress-based methods.
\black
The existing MLWMs are mostly wall-shear stress models, so we concern ourselves with wall-shear stress models only. 
The most commonly used wall-shear stress model is the equilibrium wall model \cite{deardorff1970numerical, kawai2012wall,yang2017log}, which we will use as our baseline model.
The equilibrium wall model relates the wall-shear stress with the velocity at a distance from the wall according to the law of the wall.
This incurs errors in non-equilibrium flows \cite{adler2020wall,lozano2020non,bose2018wall}, which has, on the one hand, motivated more in-depth research on the equilibrium wall model itself \cite{larsson2016large,kahraman2020adaptive,xu2021assessing,yang2018semi,chen2022wall,de2021unified,meneveau2020note} and, on the other hand, motivated research on non-equilibrium wall models \cite{park2014improved, yang2015integral, lv2021wall} and models that do not rely on the equilibrium assumption \cite{bose2014dynamic,bae2019dynamic}. 
A comprehensive review of the recent efforts on LES wall modeling is not the focus of this study, and the reader is directed to Refs. \cite{piomelli2002wall,larsson2016large,bose2018wall} as well as Ref. \cite{fowler2022lagrangian}.

The inadequacy of the equilibrium-type wall models combined with the ever-increasing availability of high-fidelity simulation data \cite{perlman2007data,graham2016web,lee2015direct} have motivated MLWMs, which is the topic of this study.
The past few years have seen the development of a number of MLWMs \cite{yang2019predictive,huang2019wall,huang2021bayesian,zhou2021wall,zhouArxiv,bae2022scientific,zhou2022wall}.
In the following, we review these efforts.
Yang \emph{et al.} \cite{yang2019predictive} appears to be the first to apply ML in WM.
They trained a feed-forward neural network (NN) against a $Re_\tau=1000$ channel to predict the LES-grid filtered wall-shear stress as a function of the LES-grid filtered velocity at a distance from the wall.
They claimed that their model predicts the mean flow in a channel at all Reynolds numbers.
Huang \emph{et al.} \cite{huang2019wall,huang2021bayesian} built on the model in Ref. \cite{yang2019predictive} and developed WMs that work well in spanwise rotating channel and channel with arbitrary (in terms of direction) but small (in terms of magnitude) system rotation.
Zhou \emph{et al.} \cite{zhou2021wall} trained a feed-forward neural network to predict the wall-shear stress as a function of the flow information at few off-wall locations.
They got good results for the periodic hill flow in their {\it a priori} tests, but their {\it a posteriori} tests did not yield good results, for which the reason was unknown.
%, which was later found to be a result of insufficient network training \cite{zhouArxiv}.
Bae \& Koumoutsakos \cite{bae2022scientific} resorted to reinforcement learning (RL).
Their method requires little knowledge of flow physics and no access to high-fidelity training data.
The trained RL model increases the wall-shear stress in response to a high velocity in the wall layer and vice versa.
The work in Ref. \cite{bae2022scientific} concerns only channel flow, and in a follow-up work, they considered flow over periodic hills \cite{zhou2022wall}.
\black Lozano \& Bae \cite{lozano2020self} developed a combined classifier and predictor network to predict wall-shear stress in the NASA juncture flow. 
Their approach yielded some improvement over EWM, but performance in separation zones remained poor despite training the model over canonical flows that include separation. 
Their method provides a useful advantage in that the classifier produces confidence levels, which can act as a warning when the MLWM's prediction is significantly outside its training range.
Bhaskaran \emph{et al.} \cite{bhaskaran2021science} developed a MLWM using unstructured compressible WRLES to predict wall-shear stress, claiming good performance in predicting complex flow boundary layers, including laminar-turbulent transition. 
However, they did not provide \emph{a posteriori} validation in this work.
Radhakrishnan \emph{et al.} \cite{radhakrishnan2021data} used gradient boosted decision trees to train a model based on a database of WRLES at $Re_\tau=180$ and $Re_\tau=1000$. 
Their \emph{a posteriori} results show predictions close to their EWM, which itself produces a log-layer mismatch. 
Moriya \emph{et al.} \cite{moriya2021inserting} used a convolutional neural network in their ML method to predict a virtual wall-surface velocity. 
Their model was trained using DNS data at $Re_\tau=180$ and was tested \emph{a posteriori} at $Re_\tau = 360$. 
However, when the grid becomes coarser ($\Delta y^+>10$), the results deteriorate. \black
%The above review is not meant to be comprehensive, and models in conferences and on arXiv are not detailed, where the authors considered junction flows, transonic compressor cascades, wall-mounted hump, and turbulent channel flow \cite{lozano2020self,bhaskaran2021science,radhakrishnan2021data,moriya2021inserting}.
 
The present work compares MLWMs from three groups, namely, the WM by Yang and co-authors \cite{yang2019predictive,huang2019wall,huang2021bayesian}, the WM by Zhou and co-authors \cite{zhou2021wall,zhouArxiv}, and the WM by Bae and co-authors \cite{bae2022scientific,zhou2022wall}.
Like any other ML papers, these models are shown to have good properties.
Yang \emph{et al.} \cite{yang2019predictive} and Zhou \emph{et al.} \cite{zhou2021wall} claimed superior results for rotating channels and periodic hills, respectively.
Bae \emph{et al.} \cite{bae2022scientific}  claimed extrapolation to unseen Reynolds numbers.
In light of the recent work by Rumsey \emph{et al.} \cite{rumsey2022search},
we will assess if these improvements are at the expense of the logarithmic law of the wall.
Along with this paper, we will make the MLWM implementation available on Github so that anyone can pick up these models as they are and apply them for predictive modeling.

Although it will be clear in the later sections, we note that problems like the log-layer mismatch (LLM) that affect empirical WMs also affect MLWMs.
LLM leads to a 15\% error in the wall-shear stress \cite{spalart2009detached}.
The error can be removed by employing a matching location away from the wall \cite{kawai2012wall}, by adding a random forcing \cite{blanchard2021stochastic}, by filtering the input velocity to the WM \cite{bou2005scale,yang2017log}, etc.
Although MLWMs have complications that prevent the application of some existing LLM remedies, this study makes all possible efforts to remove LLMs.

The rest of the paper is organized as follows. 
The computational setup and the WMs are detailed in Section \ref{sec:problem}. 
In Section \ref{sec:result}, we compare the WMLES results. 
The results are discussed in Section \ref{sec:analyse} followed by concluding remarks in Section \ref{sec:conclusion}.   

\section{\label{sec:problem} Wall-modeled large-eddy simulation details }

\subsection{\label{subsec:config} Flow configuration and code numerics}

The configuration is the half-channel flow.
The domain size is $L_x\times L_y\times L_z = 2\pi \delta \times 2\pi \delta \times 1\delta$ in the streamwise $x$, spanwise $y$, and wall-normal $z$ directions, where $\delta$ is the half channel height.
The flow is periodic in both the streamwise and the spanwise directions.
A wall-shear stress boundary condition is imposed at $z=0$, and a symmetric condition is imposed at $z=\delta$.
The flow is driven by a constant pressure gradient in the $x$ direction. 

We employ the open-source pseudo-spectral code LESGO, publicly available at \url{https://lesgo.me.jhu.edu} \cite{lesgo}.
The code uses the spectral method in the $x$ and $y$ directions and the second-order finite difference method in the $z$ direction.
The computational domain is divided uniformly into $N_x$, $N_y$, and $N_z$ grid points with the resolution of $dx$, $dy$, and $dz$ in
the $x$, $y$, and $z$ directions, respectively.
The grid planes are staggered in the vertical direction, with the first horizontal velocity plane at a distance $dz/2$ from the surface. 
The last grid point is just above the physical domain, and therefore $N_z$ grid points translate to a wall-normal grid spacing of $L_z/(N_z-1)$.
The code has been well validated and extensively used in earlier research publications \cite{albertson1999surface,porte2000scale,abkar2012new,giometto2017three,yang2018hierarchical2,yang2020scaling,yang2022logarithmic}.
Furthermore, it has served as a ground for testing SGS models and wall models \cite{bou2005scale,stoll2006dynamic,yang2015integral,abkar2016minimum,abkar2017large}. %moeng1984large,
Available SGS models include the constant coefficient \cite{smagorinsky1963general}, dynamic \cite{germano1991dynamic}, and Lagrangian dynamic \cite{meneveau1996lagrangian} Smagorinsky models, and the minimum dissipation model (AMD) \cite{rozema2015minimum,abkar2016minimum}.
Available wall models include the equilibrium wall model \cite{bou2005scale,yang2017log}, the integral wall model \cite{yang2015integral}, the slip wall model \cite{yang2016physics}, the POD-inspired wall model \cite{hansen2023pod}, and as a result of this comparative study, the supervised MLWMs in Refs. \cite{huang2019wall,zhou2021wall} and the reinforcement learning WM in Ref. \cite{bae2022scientific}.

\subsection{\label{subsec:WM} Wall models}

Four wall models are considered, namely, the equilibrium wall model, referred to as EWM \cite{moeng1984large}, the supervised MLWM in Ref. \cite{huang2019wall}, referred to as HYK19, the supervised MLWM in Refs. \cite{zhou2021wall,zhouArxiv}, denoted as ZHY21 and ZYZY22, and the reinforcement learning WM in Ref. \cite{bae2022scientific}, referred to as BK22.

\subsubsection{Empirical WM, EWM}
\label{subsubsec:EWM}

The EWM imposes the following law of the wall locally and instantaneously:
\begin{equation}
\label{eq:log_law}
    \bar{u}^+ = \frac{1}{\kappa} \ln \left(\frac{z}{z_0}\right),
\end{equation}
where $u^+=u/u_\tau$ is the inner scaled streamwise velocity, $\kappa \approx 0.4$ is the von Kármán constant, $z_0=\nu \exp(-\kappa B)/u_\tau$ is a viscous scale, and $B\approx 5$ is the intercept of the log law.
The model reads:
\begin{equation}
    \tau_w=\rho u_\tau^2=\rho\left[\frac{\kappa \tilde{U}_{\rm LES}}{\ln(h_{wm}/z_0)}\right]^2,
    \label{eq:EWM}
\end{equation}
where $\rho$ is the fluid density, $U_{\rm LES}$ is the LES horizontal velocity at a distance $h_{wm}$ from the wall, and $ \tilde{\left( \cdot \right)}$ denotes possible filtration operation \cite{yang2017log}.
Equation \ref{eq:EWM} is implicit and must be solved iteratively.
The matching height $h_{wm}$ can be the first, second, or the n\textsuperscript{th} off-wall grid point. 
In this work, we place $h_{wm}$ at $dz/2$, i.e., the first off-wall grid point, and filter the LES velocity to remove LLM \cite{yang2017log}.
Here, the factor $1/2$ is due to the use of a staggered grid.

\subsubsection{Supervised MLWM, HYK19}

\begin{figure}
\includegraphics[width=0.65\textwidth]{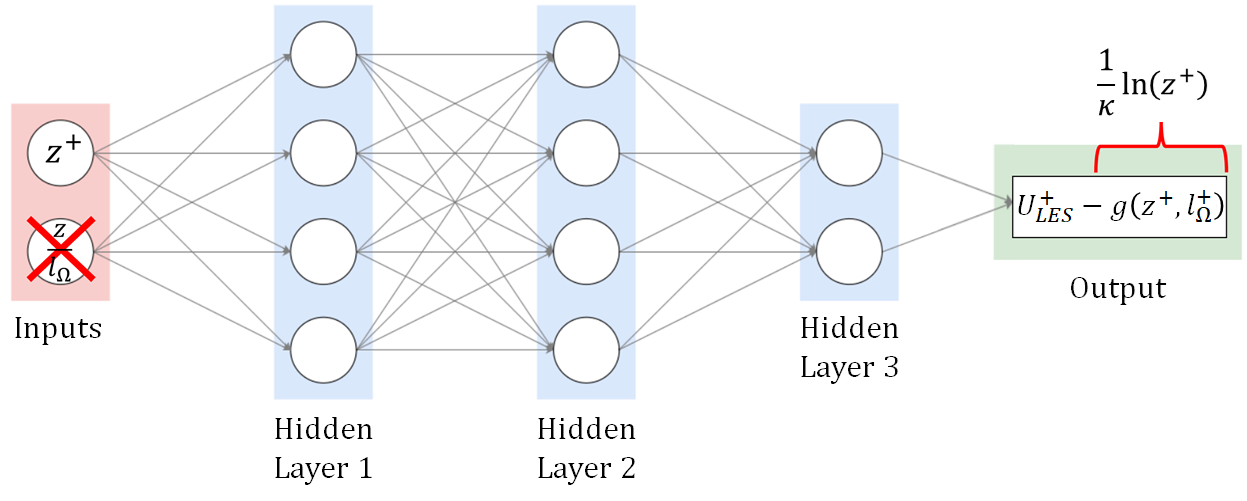}
\caption{\label{fig:schema_XRWM} Diagram representing the NN in Ref. \cite{huang2019wall}. 
Without rotation effects, the second feature is 0 and the $g$ function reduces to $\ln(z^+)/\kappa$.}
\end{figure}

It should be clear from Section \ref{subsubsec:EWM} that inverting the mean flow scaling gives a wall model.
Following that line of thinking, Huang \emph{et al.} \cite{huang2019wall} invoked the empirical knowledge in Ref. \cite{yang2020mean} and trained a network for spanwise rotating channel.
The network inputs are:
\begin{equation}
    z^+~~\text{and}~~z^+ / l_\Omega^+,
\end{equation}
where $z^+=z u_\tau / \nu$ is the wall-normal coordinate in inner unit, and $l_\Omega^+=u_{\tau}/(2\Omega)$ is the non-dimensional rotation length scale, and $\Omega$ is the angular velocity. 
In the absence of system rotation, $\Omega$ is zero, and therefore the second input feature $z^+ / l_\Omega^+$ is zero.
The network output is:
\begin{equation}
\label{eq:output}
\begin{split}
    U_{\rm LES}^+ - g(z^+,l_\Omega^+).
    \end{split}
\end{equation}
\black
The $g$ function is an empirical scaling estimate of the mean flow in a spanwise rotation channel \cite{huang2019wall,yang2020mean} defined as:
\begin{equation}
     g(z^+,l_\Omega^+) = \frac{1}{\kappa}\ln(z^+) H\left(-z^+ + \frac{l_\Omega^+}{\kappa}\right) + \left[\frac{z^+}{l_\Omega^+} + \frac{1}{\kappa} \ln(l_\Omega^+) -\frac{1}{\kappa}\ln(\kappa e) \right] H\left(z^+ - \frac{l_\Omega^+}{\kappa}\right),
\end{equation}
where $e$ is the base of the natural logarithm, $l_\Omega=u_{\tau}/(2\Omega)$ is the rotation induced length scale ($\Omega$ is the angular velocity, that drops to zero without rotation), and $H$ is the Heaviside function defined as:
\begin{equation}
    H(x) =
    \begin{cases}
      1 & \text{if $x\ge0$},\\
      0 & \text{if $x<0$}.
    \end{cases}  
\end{equation}
\black
In the absence of system rotation, the $g$ function reduces to $\ln(z^+)/\kappa$.
In that case, the network should predict the intercept of the log law at $z^+$, which varies as a function of $z$. 
\black
Huang \emph{et al.} \cite{huang2019wall} trained their network against DNS at various Reynolds numbers, ranging from $Re_\tau=180$ to $Re_\tau = 1500$, and with different rotation numbers.
\black
The network comprises three hidden layers with four, four, and two neurons in each hidden layer, as sketched in Figure \ref{fig:schema_XRWM}.
\black
Huang \emph{et al.} \cite{huang2019wall} pointed out that a small network size is necessary for computational efficiency, and larger network sizes may not be practical.
It is important to note that the neural network does not provide the wall-shear stress directly. 
Instead, an iterative process is required to deduce $u_\tau$ from the output of the network.
This is because the output depends on $U^+_{\rm LES}$ and the $g$ function (Equation \ref{eq:output}), both of which are functions of $u_\tau$. 
After several iterations, the system converges, and $u_\tau$ can be deduced from $U^+_{\rm LES}$, knowing $U_{\rm LES}$.
\black

\subsubsection{Supervised MLWM, ZHY21 and ZYZY22}

Zhou \emph{et al.} \cite{zhou2021wall} recognized the non-locality of the flow separation phenomenon and argued that it would be very hard for a wall model to accurately predict flow separation if it takes inputs at only one off-wall location.
Therefore, they took LES information at three off-wall locations and fed the information to a feed-forward neural network, which predicts the wall-shear stress.
Their network assumes information at $h_{wm}$, $h_{wm}+0.03\delta$, and $h_{wm}+0.06\delta$. 
That is, the location of the first off-wall point is arbitrary, but the second and the third off-wall points must be at distances $0.03\delta$ and $0.06\delta$ from the first off-wall point.
Table \ref{tab:ZHY_param} lists the network inputs, which contain velocity, pressure, viscosity, and coordinate information.
Their network contains 6 hidden layers, with 20 neurons in each hidden layer. 
The training is done against high-fidelity channel and periodic-hill flow data at $Re_h=5600$ to $10595$, where $Re_h$ is the Reynolds number based on the hill height.
The trained networks give superior results in {\it a priori} tests, but the {\it a posteriori} results are mixed.
\black
In recent work, Zhou \emph{et al.} \cite{zhouArxiv} re-trained their model, employing a different activation function and including channel flow data generated from the law of the wall for $Re_\tau=10^3$ to $Re_\tau=10^9$.
\black
The inputs and the outputs are also tabulated in Table \ref{tab:ZHY_param}.
\black
It should be noted that some of the features chosen in this study depend on a characteristic height of the flow, defined as $\delta$ for the half-channel flow. 
It is important to keep in mind that defining this characteristic height may be challenging in complex geometries.
\black

\begin{table*}
\begin{ruledtabular}
\caption{\label{tab:ZHY_param} Inputs, outputs and NN structure for ZHY21 \cite{zhou2021wall} and ZYZY22 \cite{zhouArxiv}. HL stands for hidden layers. The inputs make use of the near-wall scale $y^*$ defined as $y^*=\dfrac{\nu}{u_{\tau,p}}$, where $u_{\tau,p}=\sqrt{u_v^2+u_p^2}$, $u_v=\sqrt{\left|\dfrac{\nu u}{h_{wm}}\right|}$, $u_p=\left|\dfrac{\nu}{\rho}\dfrac{\partial p'}{\partial x}\right|^{1/3}$, $p'$ denotes pressure fluctuations.
}
\begin{tabular}{cccc}
WM & Inputs & Outputs &  NN size \\
\hline
\multirow{2}{*}{ZHY21} & \multirow{4}{*}{$\left[ \ln \left(\dfrac{h_{wm}}{y^*}\right) \: ; \: \dfrac{\delta}{h_{wm}}  \dfrac{u}{u_b} \: ; \: \dfrac{\delta}{h_{wm}}  \dfrac{w}{u_b} \: ; \: \dfrac{\delta}{h_{wm}}  \dfrac{v}{u_b} \: ; \:
\dfrac{h_{wm}}{\rho u_b^2} \dfrac{\partial p}{\partial x} \: ; \:
\dfrac{h_{wm}}{\rho u_b^2} \dfrac{\partial p}{\partial z} \right]\times 3$} &\multirow{4}{*}{$\dfrac{\tau_{xz_{wm}}}{u_b^2}$, $\dfrac{\tau_{yz_{wm}}}{u_b^2}$}  & \multirow{2}{*}{HL6-20} \\
 & & & \\

\multirow{2}{*}{ZYZY22} &  &  & \multirow{2}{*}{HL6-15} \\
 & & & \\
\end{tabular}
\end{ruledtabular}
\end{table*}

\subsubsection{Reinforcement learning WM, BK22}

\begin{figure}
\includegraphics[width=0.75\textwidth]{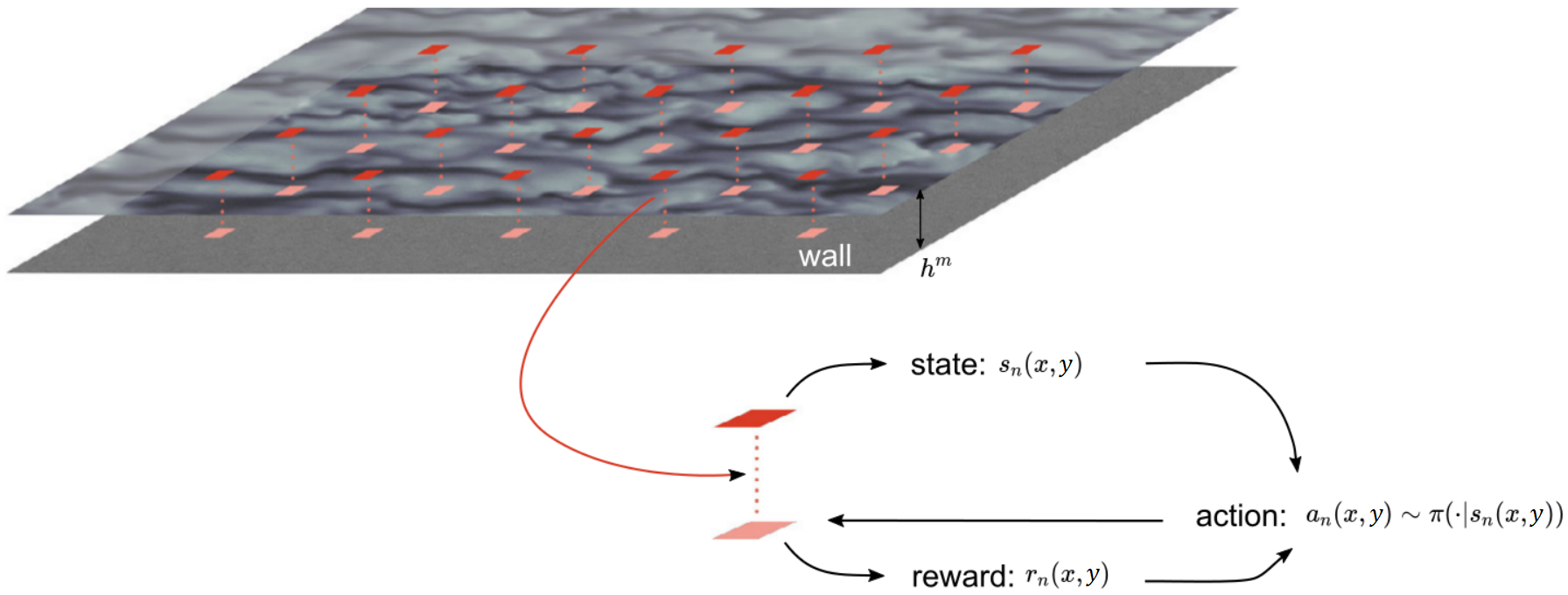}
\caption{\label{fig:schema_BWM} Several single-policy agents are distributed in the ($x,y$) plane at $h_{wm}$. Reprinted from Ref. \cite{bae2022scientific}, licensed under a Creative Commons Attribution 4.0 International License.}
\end{figure}

Bae \& Koumoutsakos \cite{bae2022scientific} resorted to reinforcement learning (RL), which does not require high-fidelity training data.
The basic idea of RL is to adjust an agent's or agents' behaviors in an environment to yield desired outcomes.
In the context of LES WM, the WM is the agent, the LES field is the environment, and the desired outcome is an accurate wall-shear stress.
Figure \ref{fig:schema_BWM} shows how the RLWM works. 
Bae \& Koumoutsakos  \cite{bae2022scientific} began their training with EWM-generated flow fields at $Re_\tau=2000$, 4200, and 8000.
Several agents are inserted into the LES flow field, and these agents modify flow fields in order to reach the best policy $\pi$. 
Specifically, the agent produces an action $a_n(x,y)$ on its environment at the instant $t_n$ based on an observation (the states $s_n$) and a reward $r_n$, causing the environment to transition from states $s_{n}$ to states $s_{n+1}$. 
The action is to increase or decrease the predicted wall-shear stress by a factor of $a_n$ as:
\begin{equation}
    \tau_w(x,y,t_{n+1}) = a_n(x,y) \tau_w(x,y,t_n).
\end{equation}
In Ref. \cite{bae2022scientific}, $a_n(x,y) \in [0.9, 1.1]$.
%The RL model was trained in WMLES at $Re_\tau=$2000, 4200, and 8000.
It is worth noting that the agents in Figure \ref{fig:schema_BWM} are single-policy agents.
That means that from given states, all agents will predict the same action.

In Ref. \cite{bae2022scientific}, two models are trained: one called the velocity wall model (VWM), and one called the log-law wall model (LLWM).
We consider only LLWM, which was shown to give better performance.
The states of LLWM are:
\begin{subequations}
    \begin{equation}
         \frac{1}{\kappa_{wm}}=\frac{h_{wm}}{u_{\tau_{wm}}} \left.\frac{\partial u}{\partial z}\right|_{z=h_{wm}}, 
    \end{equation}
    \begin{equation}
         B_{wm} = \frac{u_{\rm LES}}{u_{\tau_{wm}}} - \frac{1}{\kappa_{wm}} \ln \left( \frac{h_{wm} u_{\tau_{wm}}}{\nu} \right),
    \end{equation}
\end{subequations}
where $u_{\rm LES} = u(h_{wm})$ is the streamwise LES velocity taken at $h_{wm}$ and $u_{\tau_{wm}}=\sqrt{\tau_w/\rho}$ is the model-computed friction velocity. 
We place the matching location $h_{wm}$ at $dz$, i.e., between the first and the second off-wall grid points, for ease of computing the velocity derivative.
It may be worth noting that the implementation of RLWM is not straightforward. 
The present model requires the coupling of an RL library, here the Smarties library \cite{novati2019a} with the LES solver.
Ideally, the trained model should not need the RL library, a topic we will leave to future investigation.

\subsection{Further details}

Table \ref{tab:overview} summarizes the WMLES cases considered in this study.
We vary the grid resolutions from $N_x \times N_y \times N_z = 24^3$ to $N_x \times N_y \times N_z = 72^3$.
The sub-grid scales are modeled with the AMD model \cite{rozema2015minimum,abkar2016minimum}.
\black
We have evaluated an additional sub-grid scale (SGS) model, the Lagrangian scale-dependent dynamic (LSD) SGS model \cite{bou2005scale}, at $Re_\tau=5200$, and the corresponding results are presented in Appendix \ref{app:DS}.
\black
Eleven friction Reynolds numbers are considered, between $180$ and $10^{10}$. 
For ZHY21 and ZYZY22, the distances between the first and the second off-wall locations must equal the distance between the second and the third off-wall locations, which must equal $0.03\delta$.
This limits the wall-normal grid resolution to about $dz=0.03\delta$ unless one interpolates, which will likely introduce unwanted errors.
This is why the grid size is $32^3$ for ZHY21 and ZYZY22.
\black It should be noted that the matching point location for BK22 differs from other WMs due to the computation of derivatives in the definition of states.
The EWM and HYK19 models differ from others in that they use filtered inputs. 
The decision not to filter inputs for other models is based on how they were originally trained and tested, which was without filtering. 
Additionally, using filtered inputs for these models requires filtering inputs at several distances from the wall, as these models require information from multiple points.

\black

\begin{table*}
\begin{ruledtabular}
\caption{\label{tab:overview} Details of the WMLESs. \black In some cases, inputs are filtered using a sharp filter with a side length equal to twice the grid spacing. }
\begin{tabular}{ccccccc}
Wall model & SGS model & \makecell{Grid size \\$\left(N_x \times N_y \times N_z\right)$} &  $Re_\tau$ & $N_\textrm{agents}$ & \black \makecell{Matching \\location ($h_{wm}$)} & \black \makecell{Inputs \\ filtering}  \\
\hline
{EWM} & AMD & $24^3$, $48^3$, $72^3$  & {[180, $10^{10}$]} & {/} &\black $dz/2$  &\black Yes \\
{HYK19} & AMD & $24^3$, $48^3$, $72^3$  & {[180, $10^{10}$]} & {/} &\black $dz/2$ &\black Yes\\
ZHY21, ZYZY22 & AMD & $32^3$  & [180, $10^{10}$] & / &\black $dz/2$ &\black No \\
{BK22} & AMD & $24^3$, $48^3$, $72^3$  & {[180, $10^{10}$]} &  16, 64, 128 &\black $dz$ &\black No\\
\end{tabular}
\end{ruledtabular}
\end{table*}

\section{\label{sec:result}Results}

Before we present the WMLES results, it would be instructive to mention two comments from Brenner \& Koumoutsakos \cite{brenner2021machine} and Rumsey \emph{et al.} \cite{rumsey2022search}.
Brenner \& Koumoutsakos \cite{brenner2021machine} noted that one must apply the same standards to ML works.
Rumsey \emph{et al.} \cite{rumsey2022search} noted:
``the (ML) model should be ‘universal’ in the sense that it can be used by anyone and applied to as many flows as possible without concern for unusual or detrimental behavior; at worst, the ML model should not degrade the accuracy of the baseline model.''
\black
We tested the already trained ML models provided by the authors in our solver and did not retrain them.
The performance of the ML models is affected by the differences in the CFD solver used in our testing, which highlights the need for these models to be adapted to account for such differences.
\black

\subsection{The effects of wall model}

\begin{figure}
\centering
\subfigure[EWM.]{\includegraphics[width=0.3\textwidth]{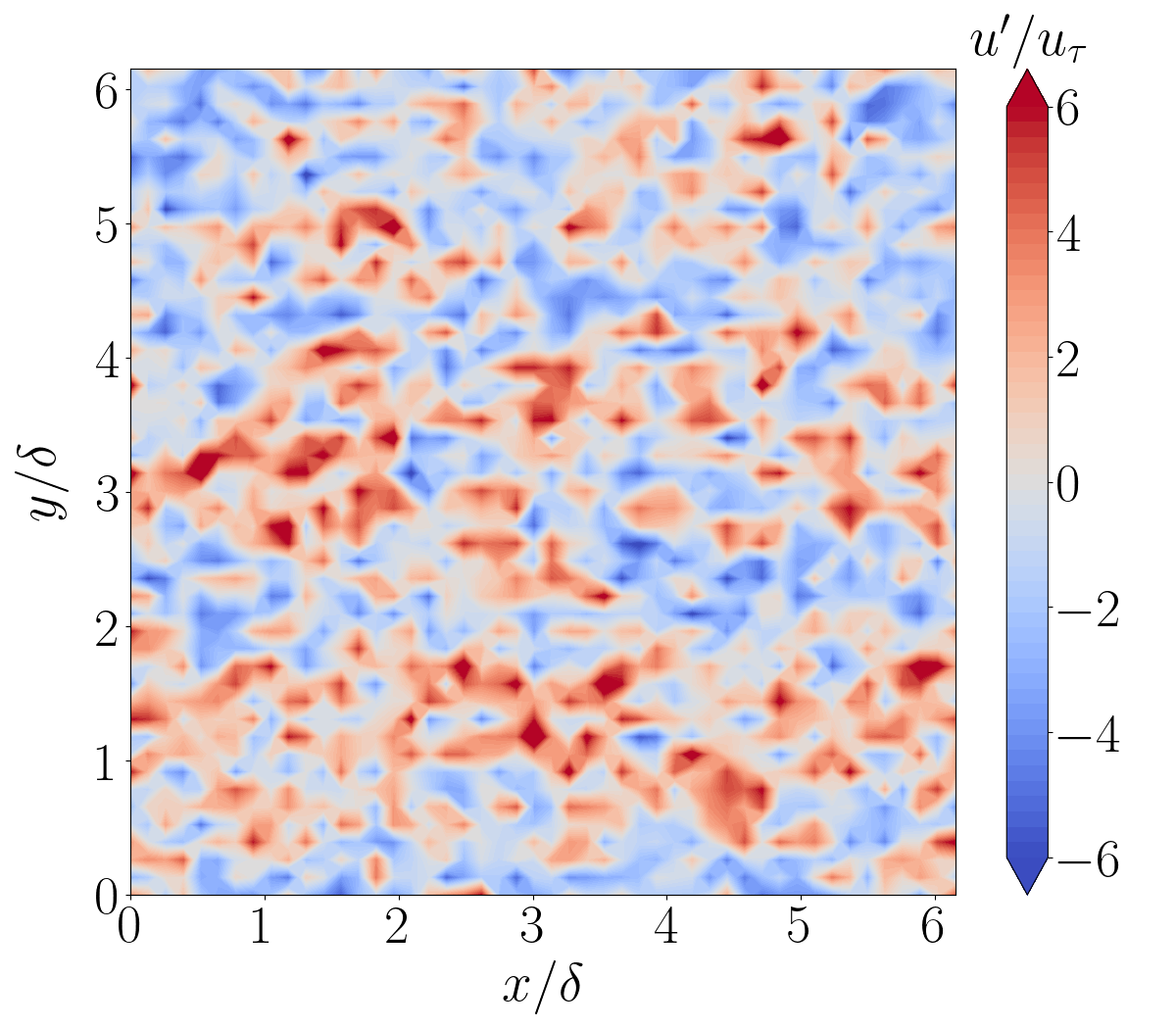}}
\subfigure[HYK19.]{\includegraphics[width=0.3\textwidth]{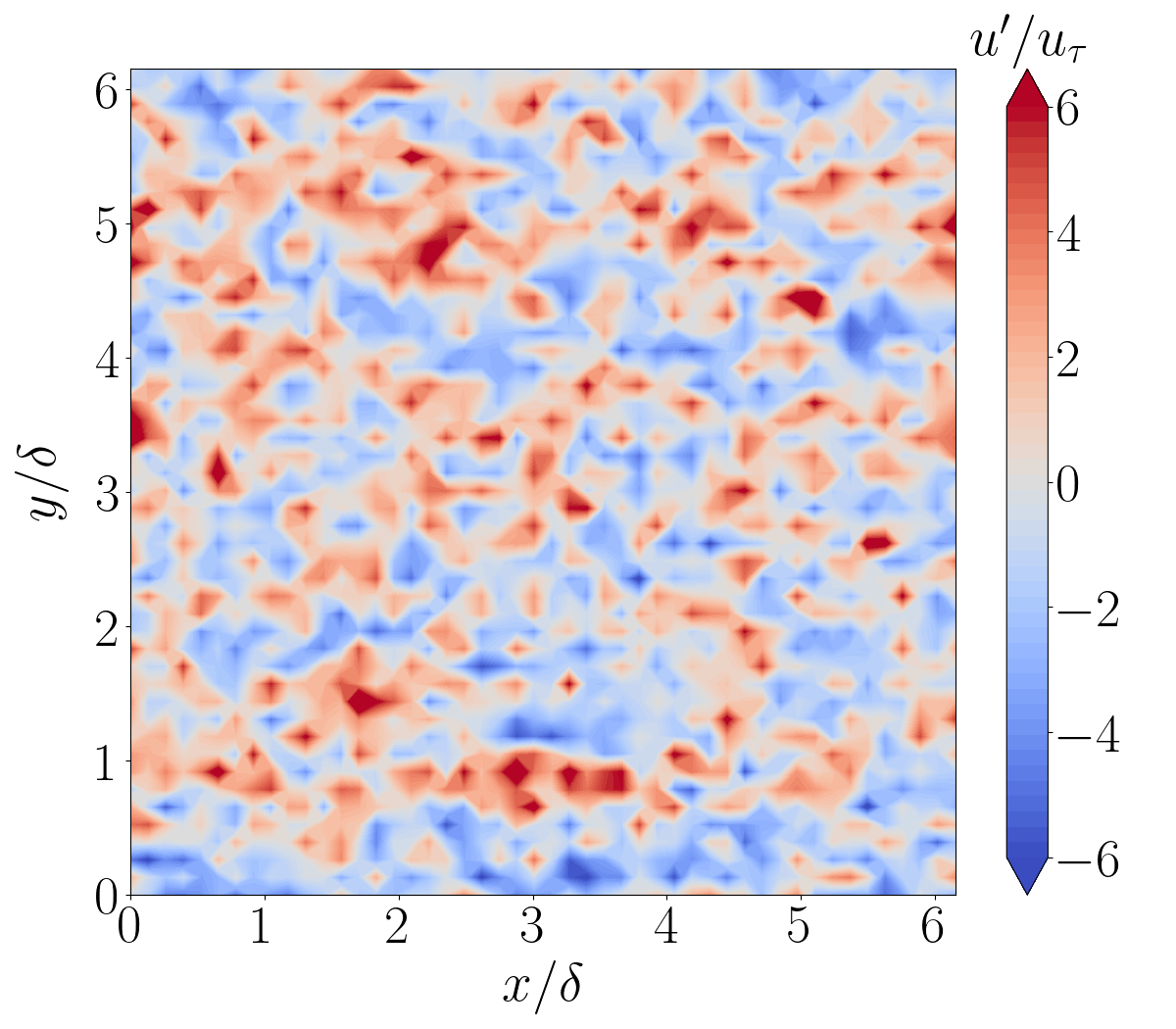}}
\subfigure[BK22.]{\includegraphics[width=0.3\textwidth]{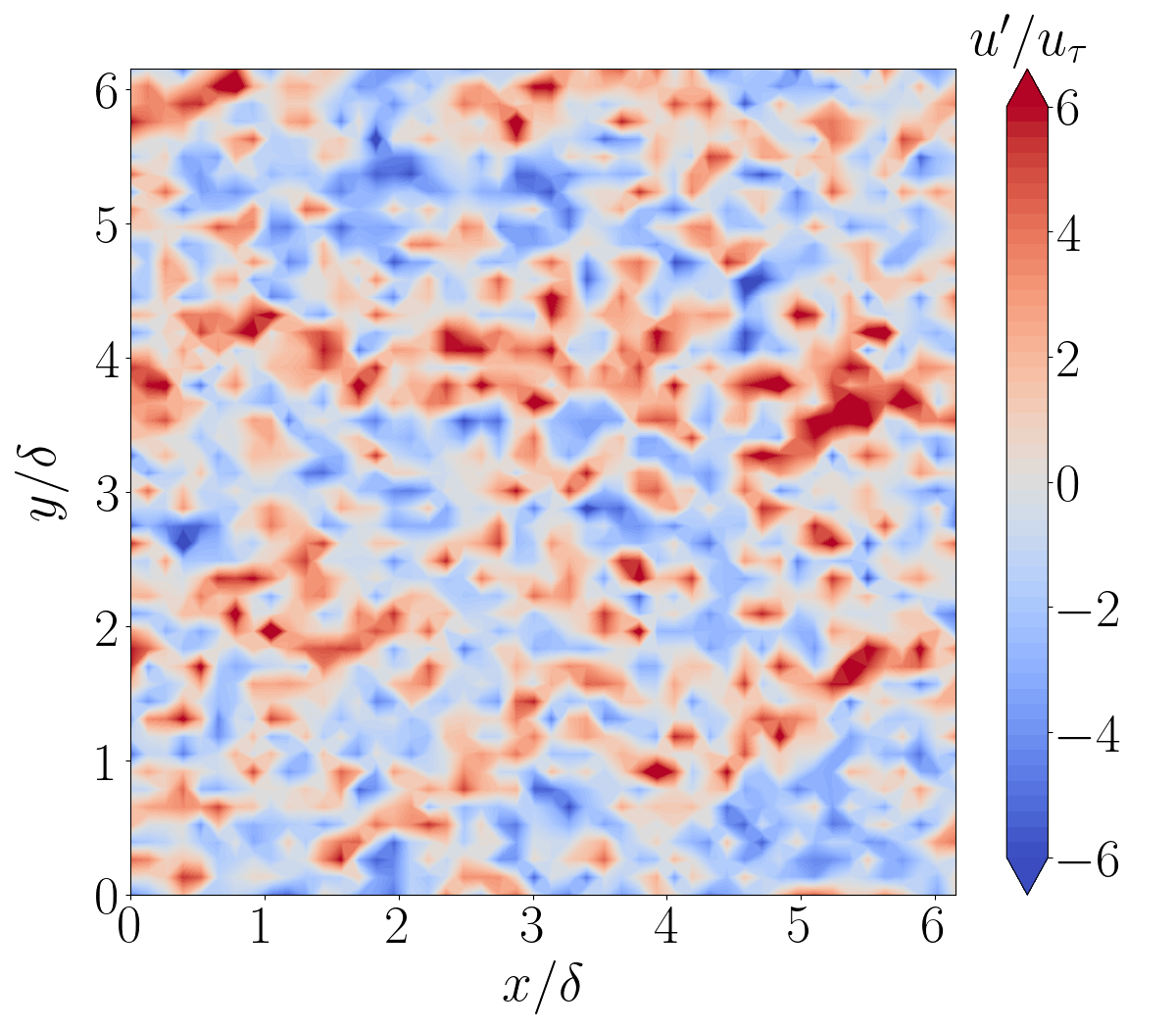}}
\caption{\label{fig:planxy1}Contours of the fluctuating streamwise velocity in the $x-y$ plane at the first off-wall grid point $z=dz/2$.
The flow is at $Re=10^5$.
(a) EWM \cite{kawai2012wall}, (b) HYK19 \cite{huang2019wall}, (c) BK22 \cite{bae2022scientific}. 
The grid resolution is $N_x \times N_y \times N_z = 48^3$.} 
\end{figure}

\begin{figure}
\centering
\subfigure[EWM.]{\includegraphics[width=0.3\textwidth]{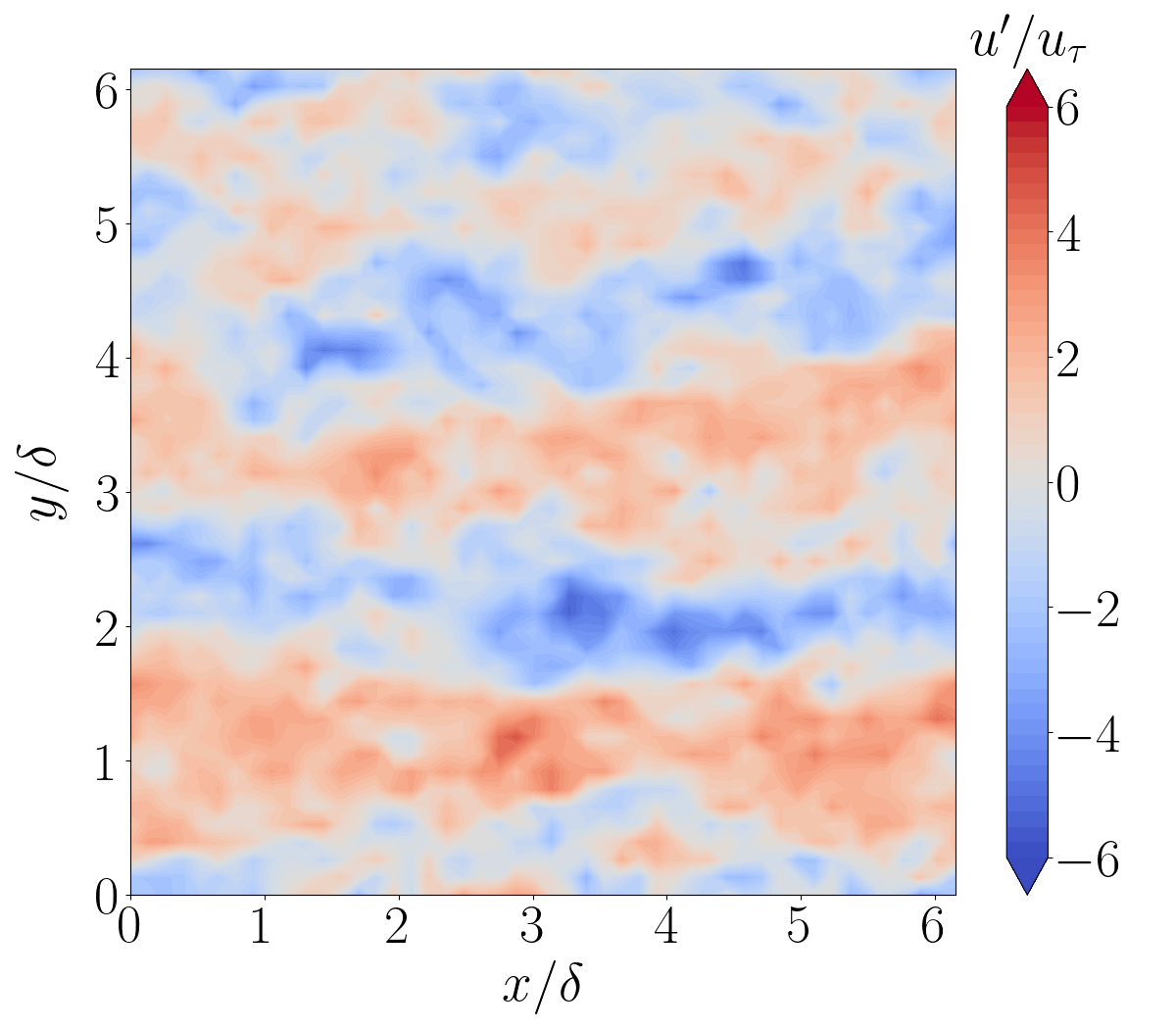}}\label{fig:planxy2_EWM}
\subfigure[HYK19.]{\includegraphics[width=0.3\textwidth]{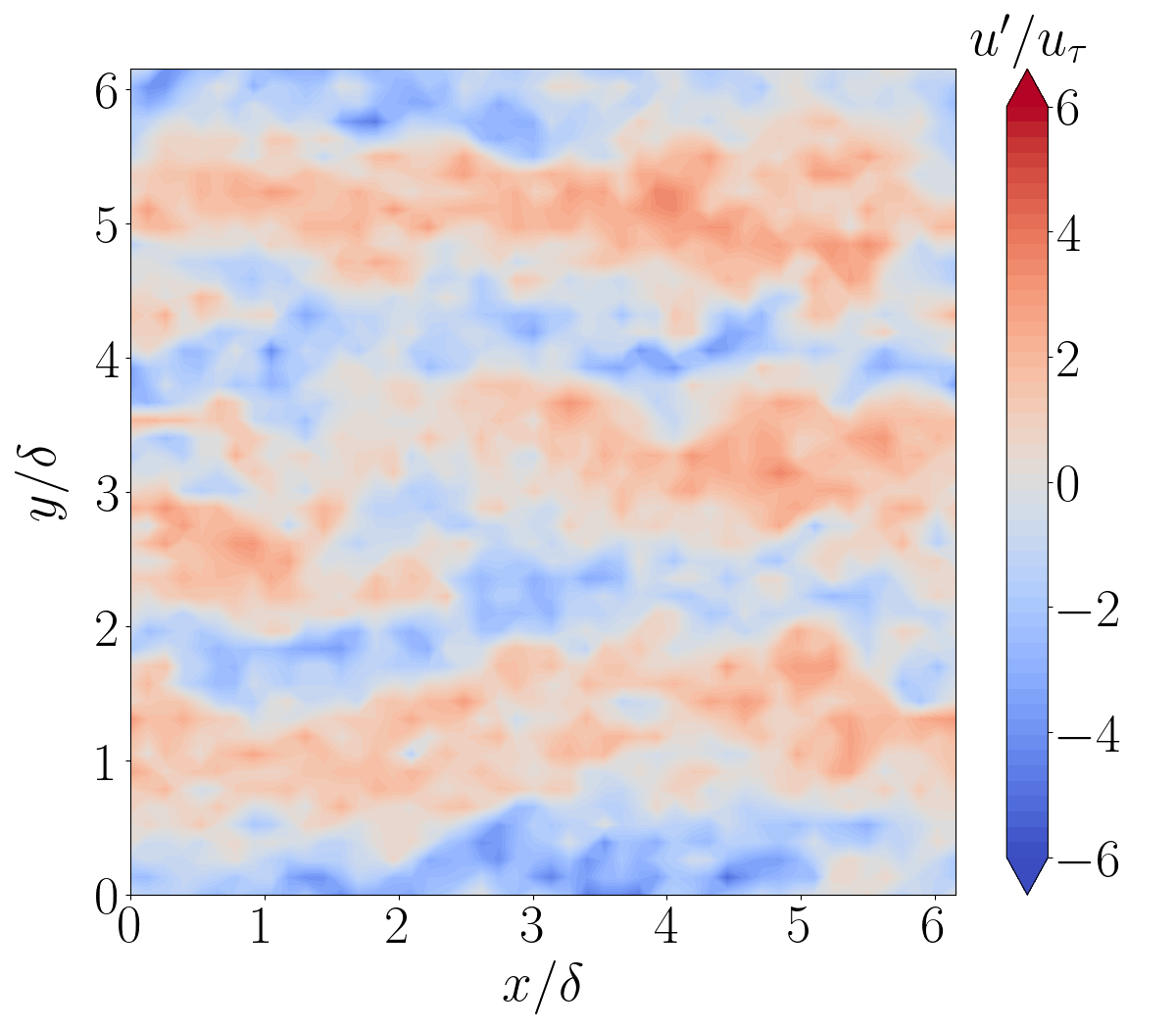}}\label{fig:planxy2_HXK}
\subfigure[BK22.]{\includegraphics[width=0.3\textwidth]{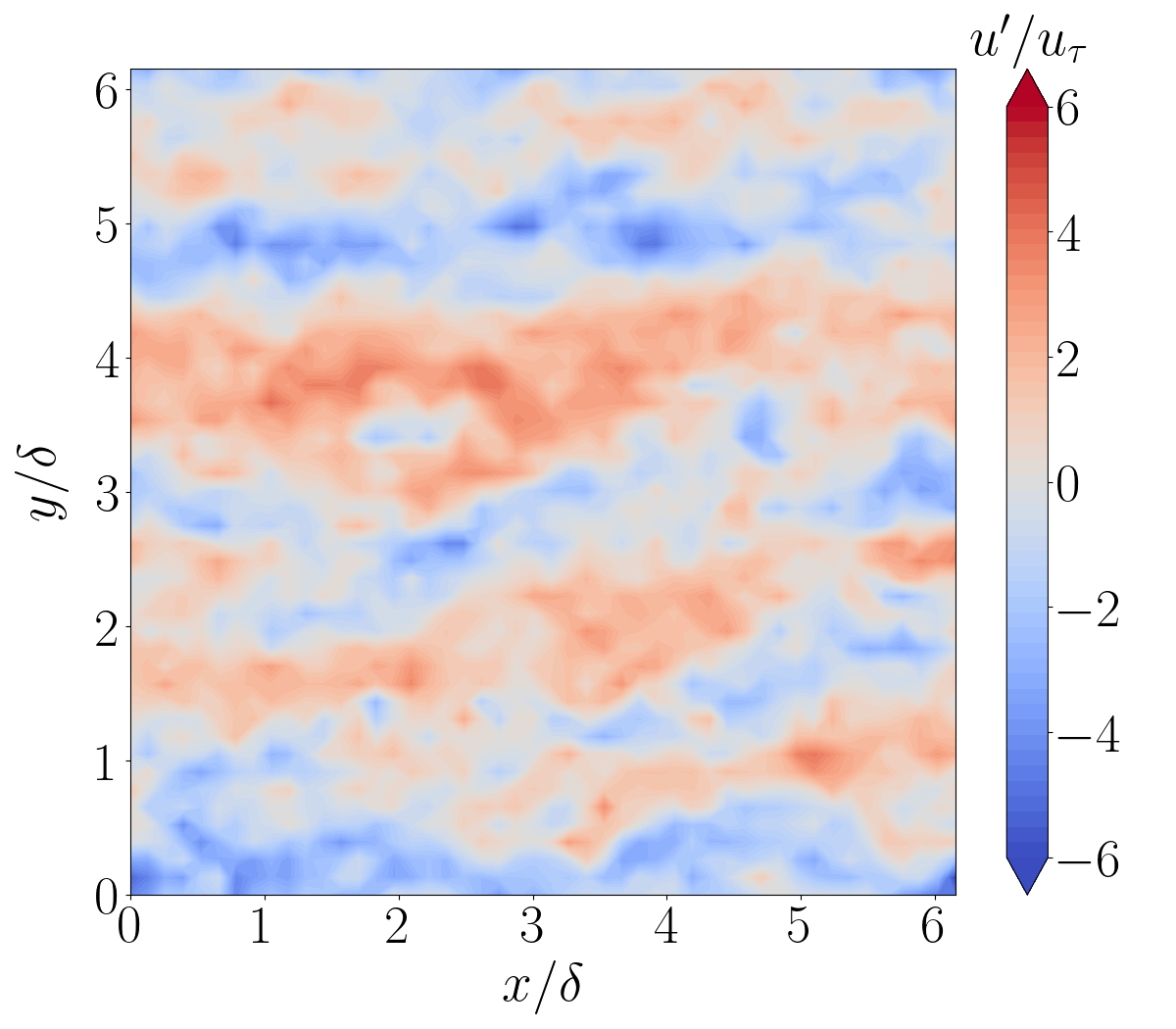}}\label{fig:planxy2_BK22}
%\subfigure[ZYZY22.]{\includegraphics[width=0.4\textwidth]{Figures/planxy2_ZWM.png}}\label{fig:planxy2_ZYZY22}
\caption{\label{fig:planxy2}Same as figure \ref{fig:planxy1} but at $z/\delta=0.5$.} 
\end{figure}
  
\begin{figure}
\centering
\subfigure[EWM.]{\includegraphics[width=0.4\textwidth]{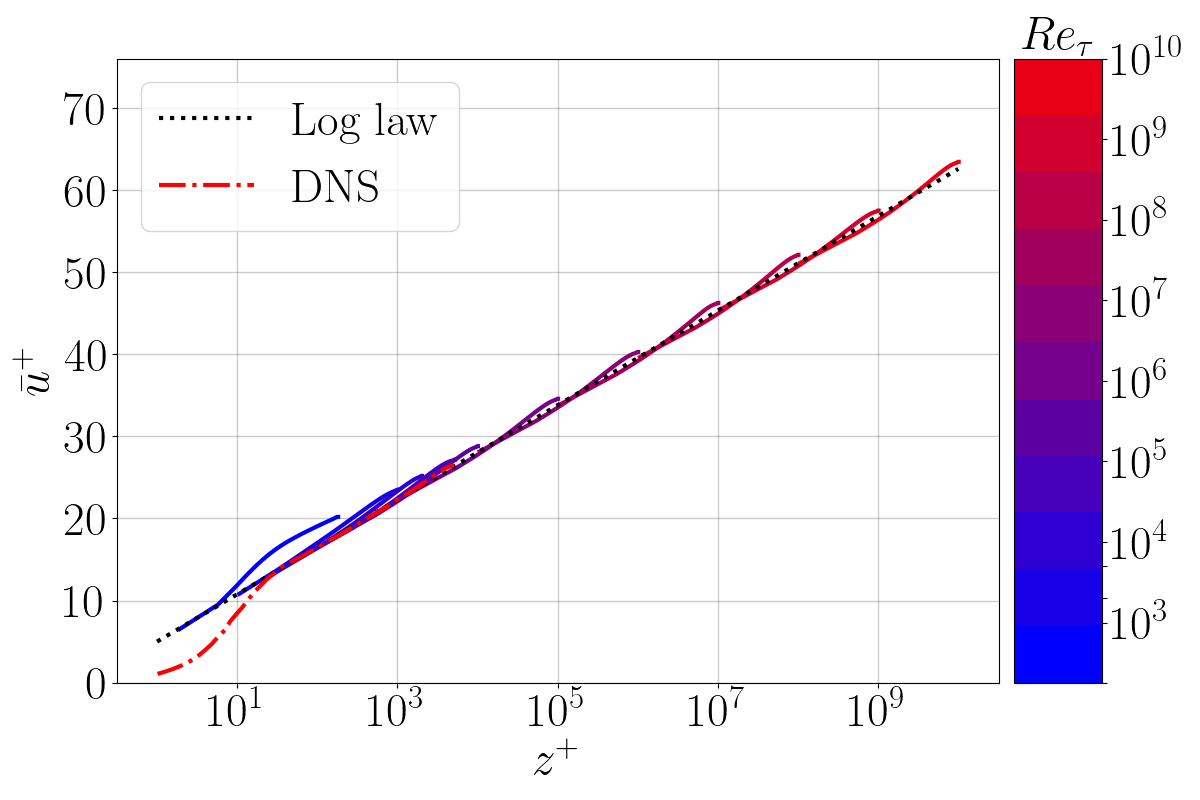}}
\subfigure[HYK19.]{\includegraphics[width=0.4\textwidth]{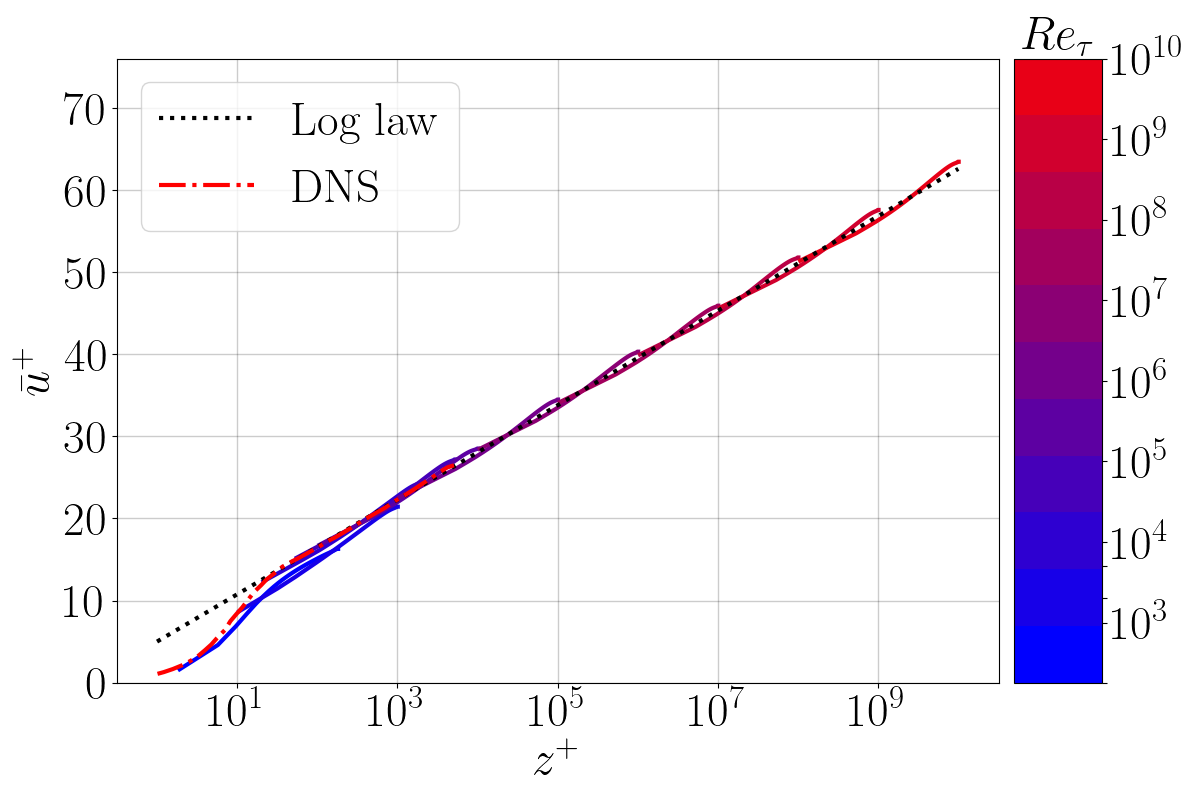}}
\subfigure[ZYZY22.]{\includegraphics[width=0.4\textwidth]{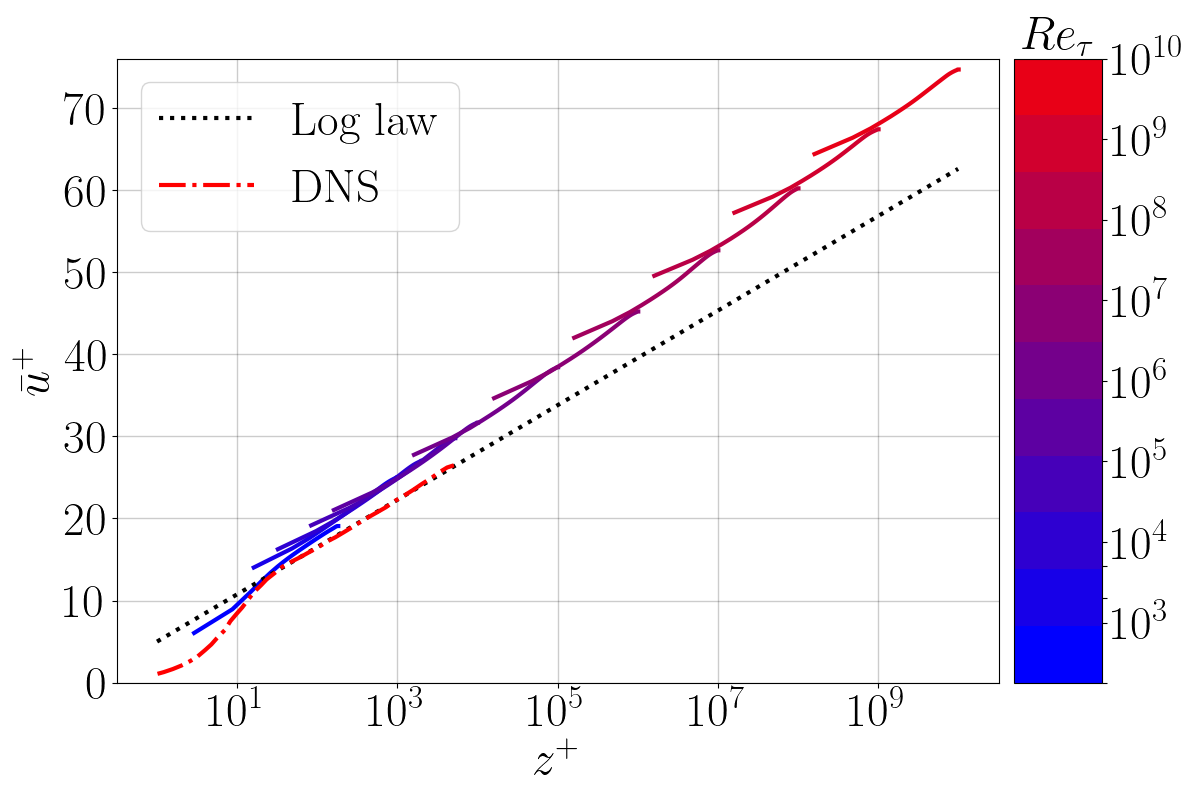}}
\subfigure[BK22.]{\includegraphics[width=0.4\textwidth]{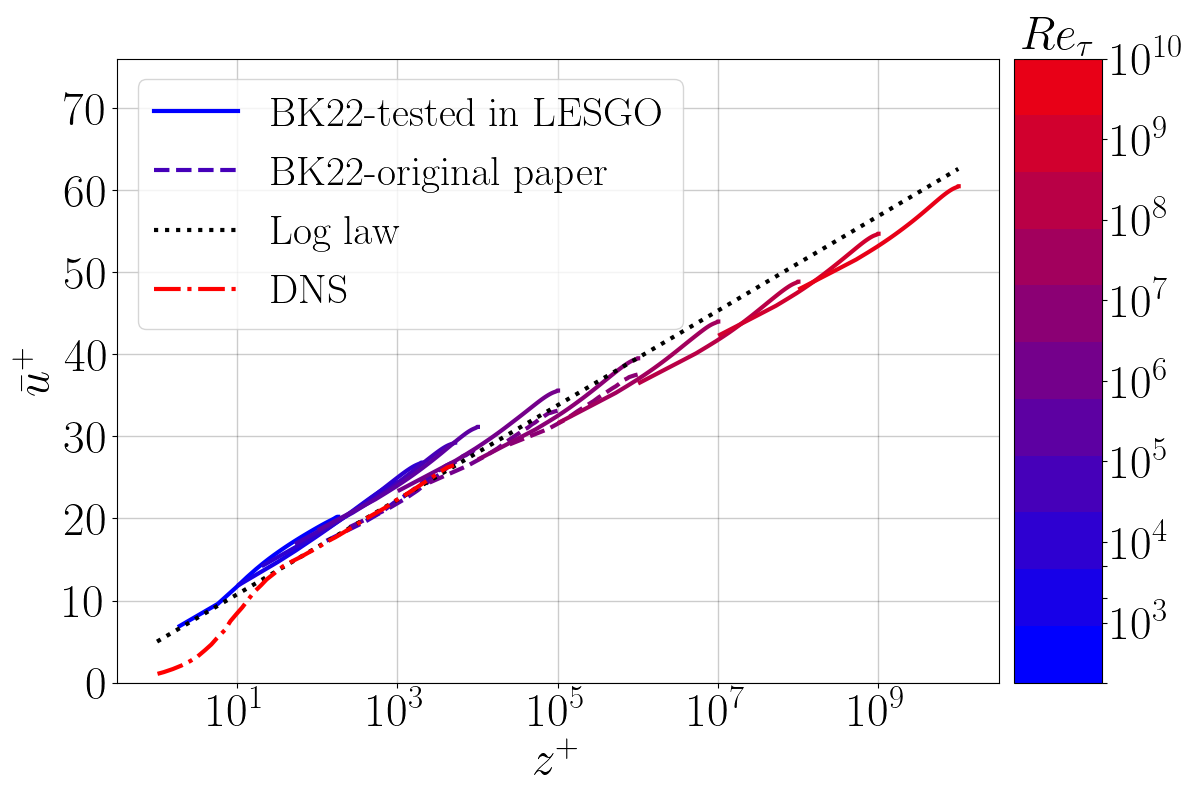}}
\caption{\label{fig:res}Mean streamwise velocity $\bar{u}^+$ as a function of the wall-normal direction $z^+$ at 11 Reynolds numbers between $Re_\tau=180$ and $10^{10}$.
(a) EWM \cite{kawai2012wall}, (b) HYK19 \cite{huang2019wall}, (c) ZYZY22 \cite{zhouArxiv}, (d) BK22 \cite{bae2022scientific}. 
DNS result at $Re_\tau=5200$ is included for comparison purposes \cite{lee2015direct}.
The log law corresponds to $\kappa=0.4$ and $B=5$.
\black The comparison of results for BK22 includes two sets, one from the original paper \cite{bae2022scientific} at $Re_\tau=5200$, $10^4$, $10^5$, and $10^6$ and the other from an implementation of the original model in LESGO.}
\end{figure}

\begin{figure}
\includegraphics[width=0.4\textwidth]{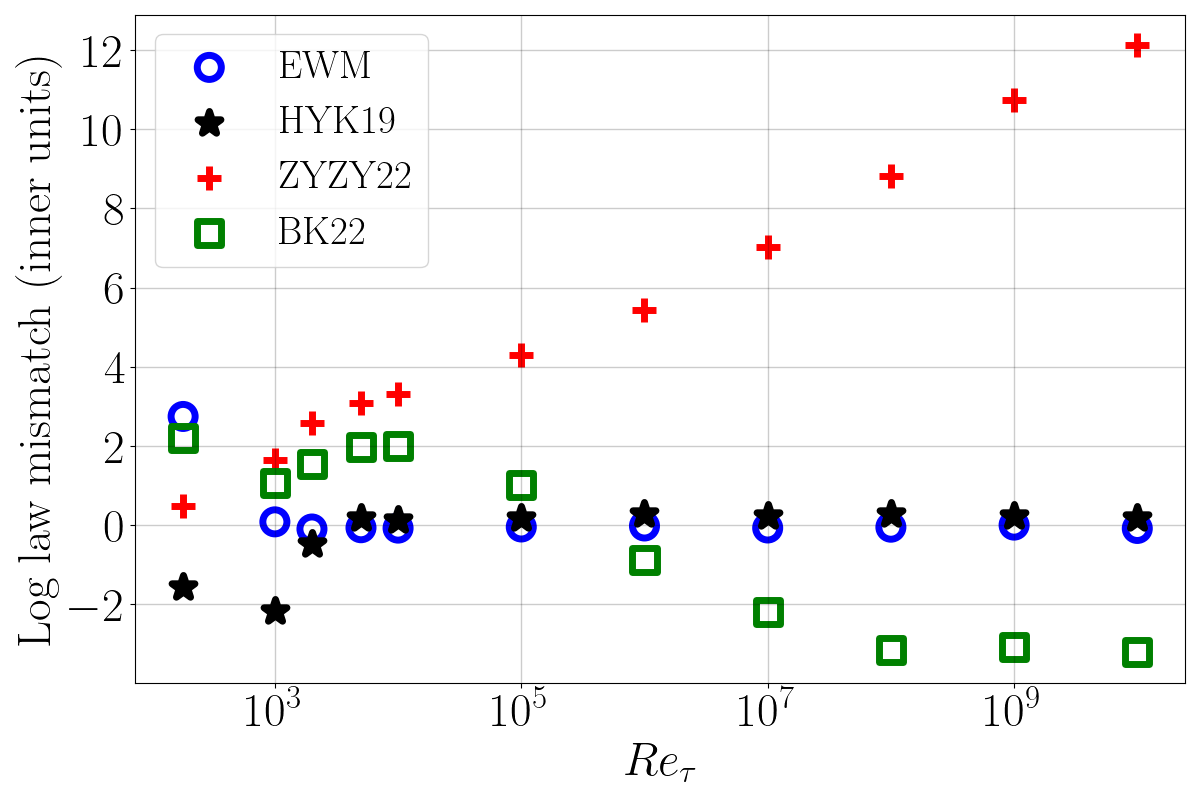}
\caption{\label{fig:mismatch_LL} LLM in inner units as a function of the friction Reynolds number $Re_\tau$ for all wall models.
The baseline log law is $\ln(z^+)/\kappa+B$ with $\kappa=0.4$ and $B=5$.}
\end{figure}

Figures \ref{fig:planxy1} and \ref{fig:planxy2} show the contours of the streamwise velocity at the first off-wall grid point and at $z/\delta=0.5$.
The Reynolds number is $Re_\tau=10^5$.
The grid resolution is $N_x\times N_y\times N_z=48^3$, and 64 agents are employed for the BK22 results.
Results of EWM, HYK19, and BK22 are shown.
ZYZY22 results are available at only the resolution $N_x\times N_y\times N_z=32^3$ and are not shown here for fairness.
The flow fields at the same height are alike:
we see small-scale large fluctuations at the first off-wall grid point and large-scale streaks at $z/\delta=0.5$ in all WMLESs irrespective of the wall model.
Figure \ref{fig:res} shows the mean velocity profiles.
The EWM captures the log law when the matching location $h_{wm}$ is in the log layer.
When the matching location is in the viscous layer, i.e., at $Re_\tau=180$, the algebraic EWM employed here yields a profile above the DNS.
HYK19 gives the right mean flow at all Reynolds numbers, irrespective of where $h_{wm}$ is.  
ZYZY22 predicts the log law but yields a small von Kármán constant---in spite of the fact that all results here are from the same SGS model.
This is quite peculiar, and it is not clear to the authors why ZYZY22 yields a small von Kármán constant.
Errors are found at $Re_\tau=180$ like the EWM.
BK22 also predicts the log law, but positive and negative LLMs are found at low and high Reynolds numbers, respectively.
Errors are also found in BK22 at $Re_\tau=180$.
\black 
Since the log law is only exact at infinite Reynolds number, a closer comparison of the mean velocity profiles with DNS is shown in Appendix \ref{app:DNS} for $Re_\tau=180$, $1000$ and $5200$.
\black
In addition, we precisely measure LLM, and the result is shown in Figure \ref{fig:mismatch_LL}.
The increase in the measured LLM as a function of the friction Reynolds number in ZYZY22 is because ZYZY22 yields a small von Kármán constant:
\begin{equation}
    {\rm LLM}\sim \left(\frac{1}{\kappa_{\rm ZYZY22}}-\frac{1}{\kappa}\right)\ln(z^+),
\end{equation}
where $\kappa_{\rm ZYZY22}$ is the ZYZY22 predicted von Kármán constant.
Bae \& Koumoutsakos reported results up to $10^6$  \cite{bae2022scientific}.
Their results are slightly different from the ones shown here mainly because of the difference in the grid resolution.
The good performance at $Re_\tau=10^5$ and $10^6$ is because the $h_{wm}^+$ value is near the values the model is trained for, and the mismatches at higher and lower Reynolds numbers are because the $h_{wm}^+$ value is far from the training conditions.
This is particularly true for $Re_\tau=180$, where $h_{wm}^+$ is inside the viscous layer and the flow there does not abide by the log law.
\black 
As shown in Appendix \ref{app:DS} for $Re_\tau=5200$, the results are not impacted by the size of the domain, which is sufficiently large.
\black

\begin{figure}
\centering
\subfigure[EWM.]{\includegraphics[width=0.4\textwidth]{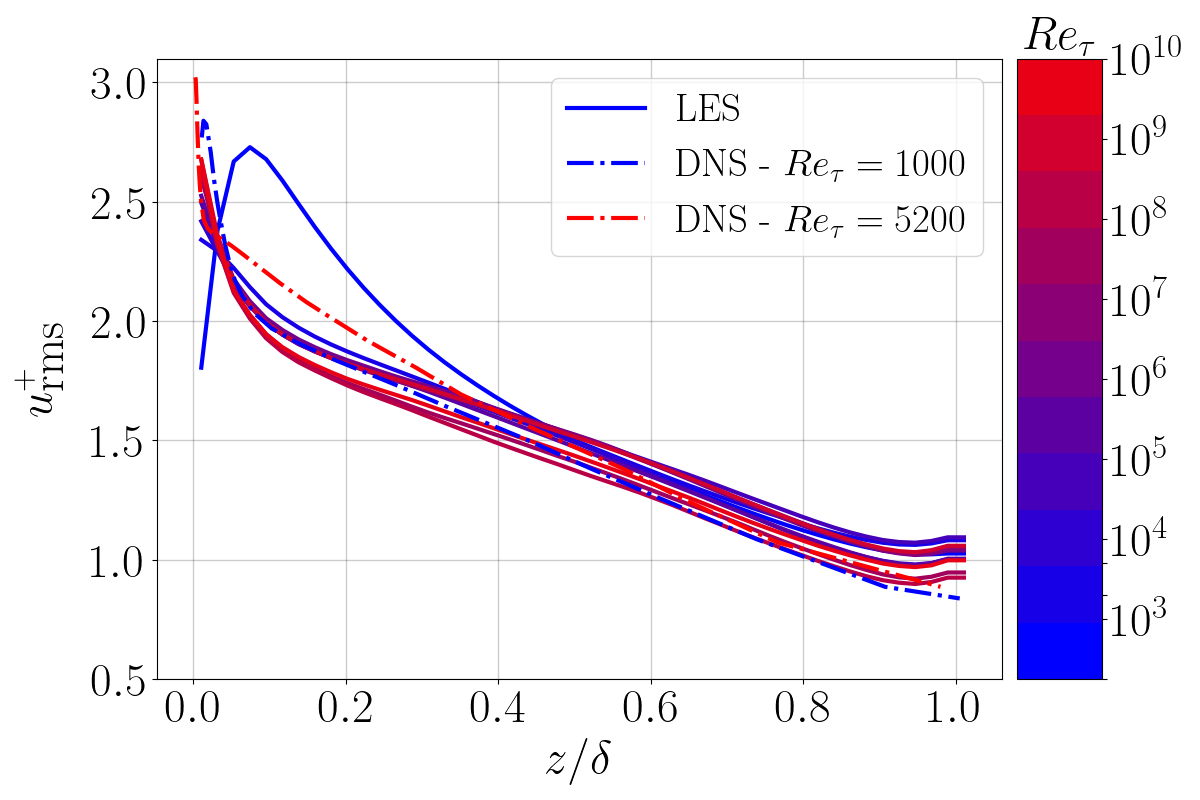}}
\subfigure[HYK19.]{\includegraphics[width=0.4\textwidth]{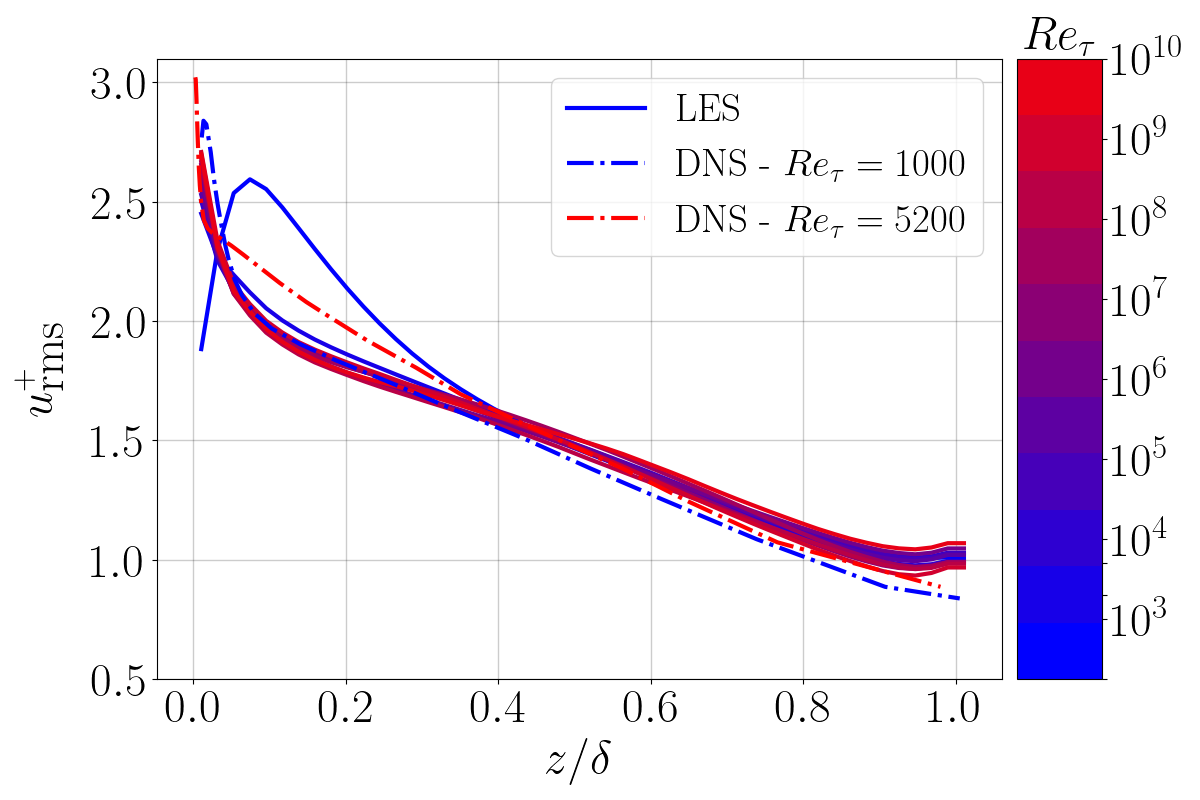}}
\subfigure[ZYZY22.]{\includegraphics[width=0.4\textwidth]{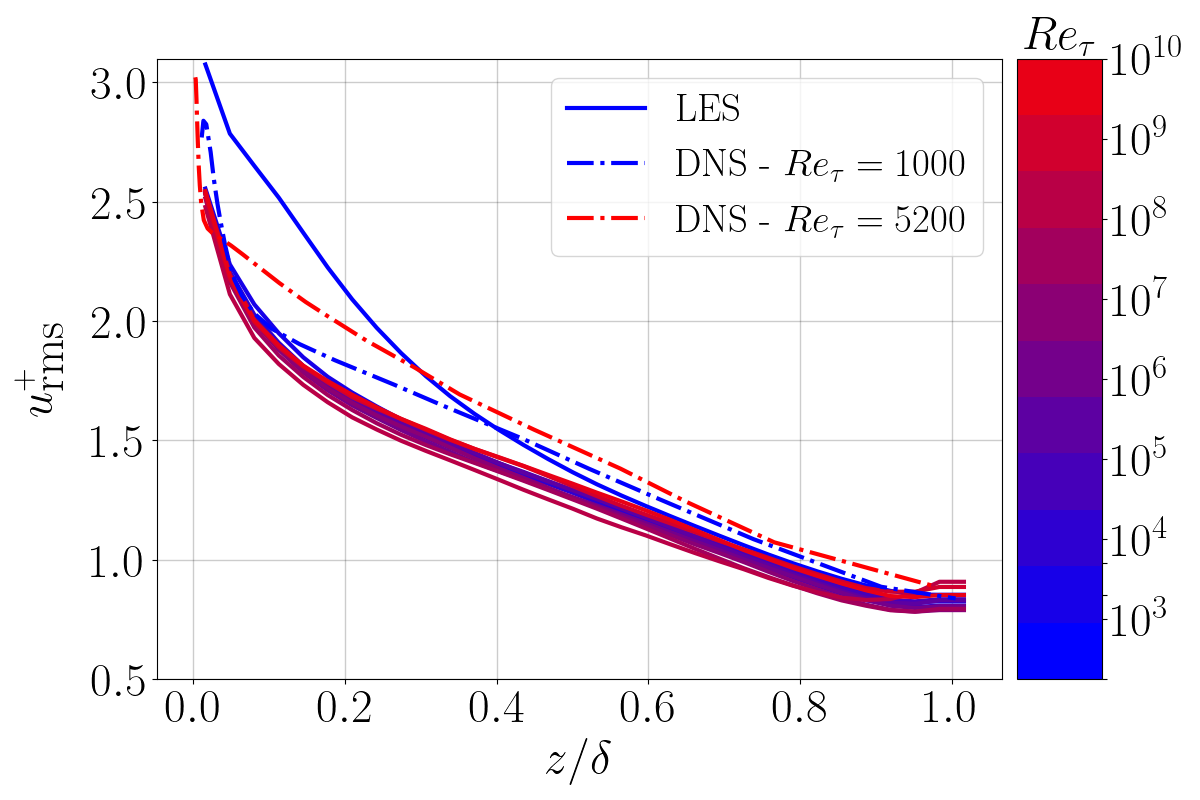}}
\subfigure[BK22.]{\includegraphics[width=0.4\textwidth]{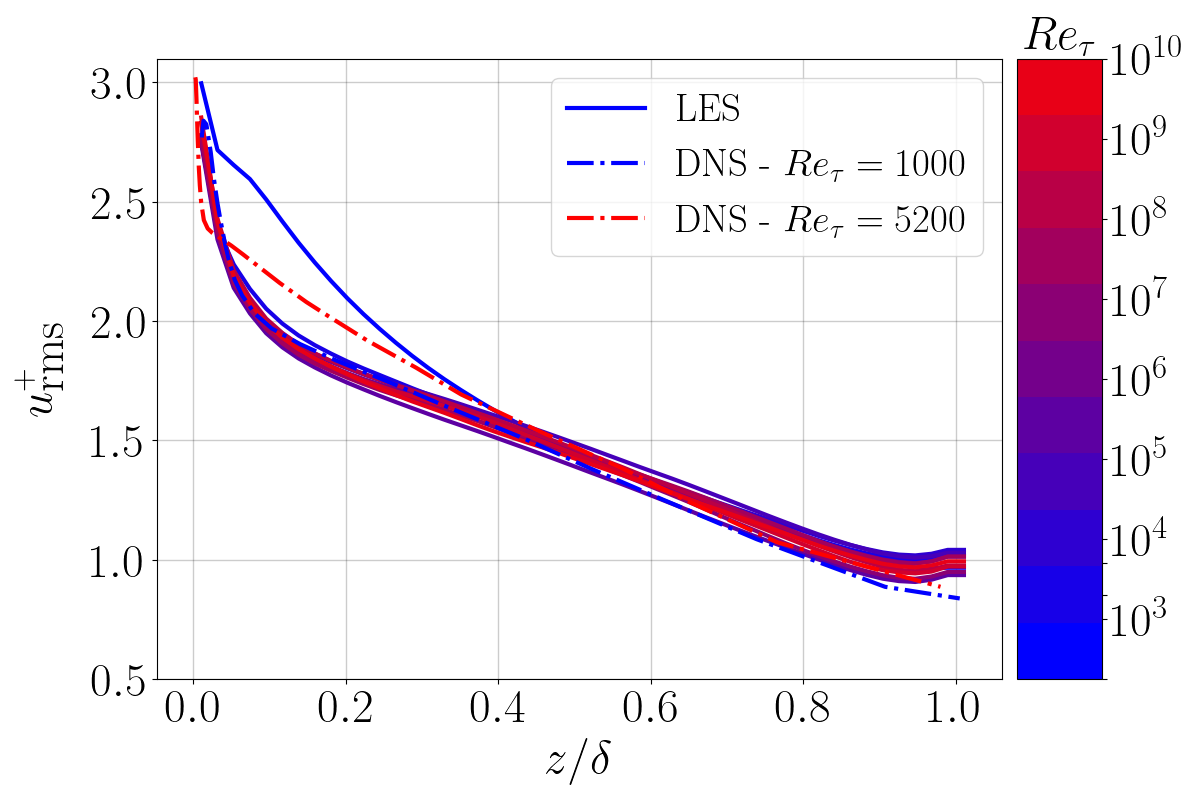}}
\caption{\label{fig:rms} Root mean square of the streamwise velocity fluctuation $u_\textrm{rms}^+$ as a function of the wall-normal coordinate $z/\delta$ plotted for $Re_\tau \in \left[180, 10^3, 2\times10^3, 5.2\times 10^3, 10^4, 10^5, 10^6, 10^7, 10^8, 10^9, 10^{10}\right]$. (a) EWM \cite{kawai2012wall}, (b) HYK19 \cite{huang2019wall}, (c) ZYZY22 \cite{zhouArxiv}, (d) BK22 \cite{bae2022scientific}. DNS data at $Re_\tau=1000$ \cite{graham2016web} and $Re_\tau=5200$ \cite{lee2015direct} are included for comparison. } 
\end{figure}

Figure \ref{fig:rms} shows the inner scaled root-mean-square of the streamwise velocity fluctuations $u_{\rm rms}^+$.
The $Re_\tau=180$ result is different from the rest due to its low Reynolds number.
The $u_{\rm rms}^+$ at a fixed $y/\delta$ should not vary as a function of the Reynolds number at sufficiently high Reynolds numbers \cite{marusic2019attached,yang2019hierarchical}.
This is what we see in Figure \ref{fig:rms}.
The EWM and HYK19 results are similar.
ZYZY22 gives a large $u_{\rm rms}^+$ at $Re_\tau=180$ but the results are otherwise similar to these of the EWM.
BK22 gives a slightly larger $u_{\rm rms}^+$ at the first off-wall grid point than the EWM, which is quite peculiar.

\subsection{Further results}

In this subsection, we report the effects of the number of agents and the grid resolution. %, and sub-grid scale model.

\begin{figure}
\centering
\subfigure[$N_\textrm{agents}= 16$.]{\includegraphics[width=0.325\textwidth]{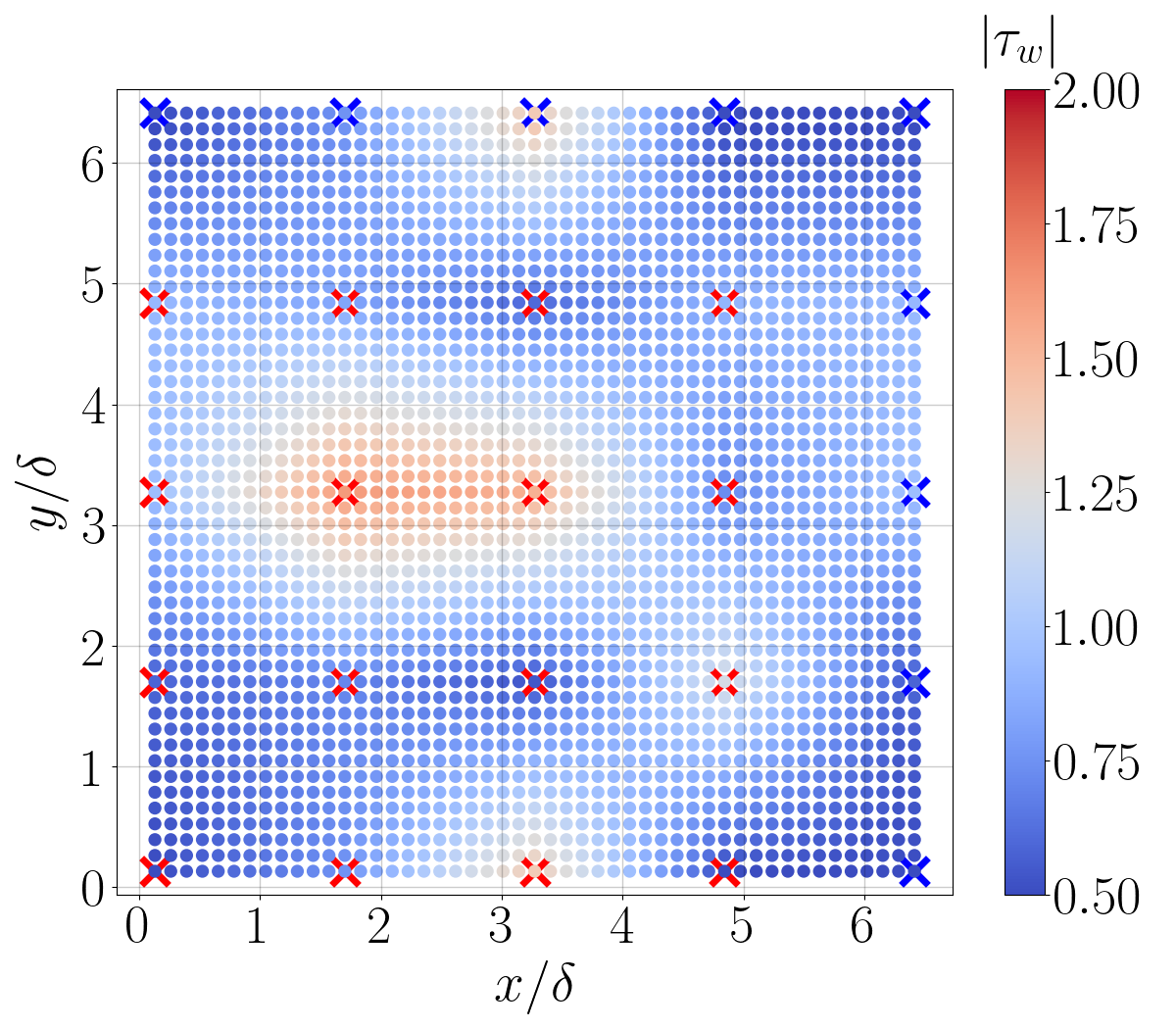}}\label{fig:Nag16}
\subfigure[$N_\textrm{agents}= 64$.]{\includegraphics[width=0.325\textwidth]{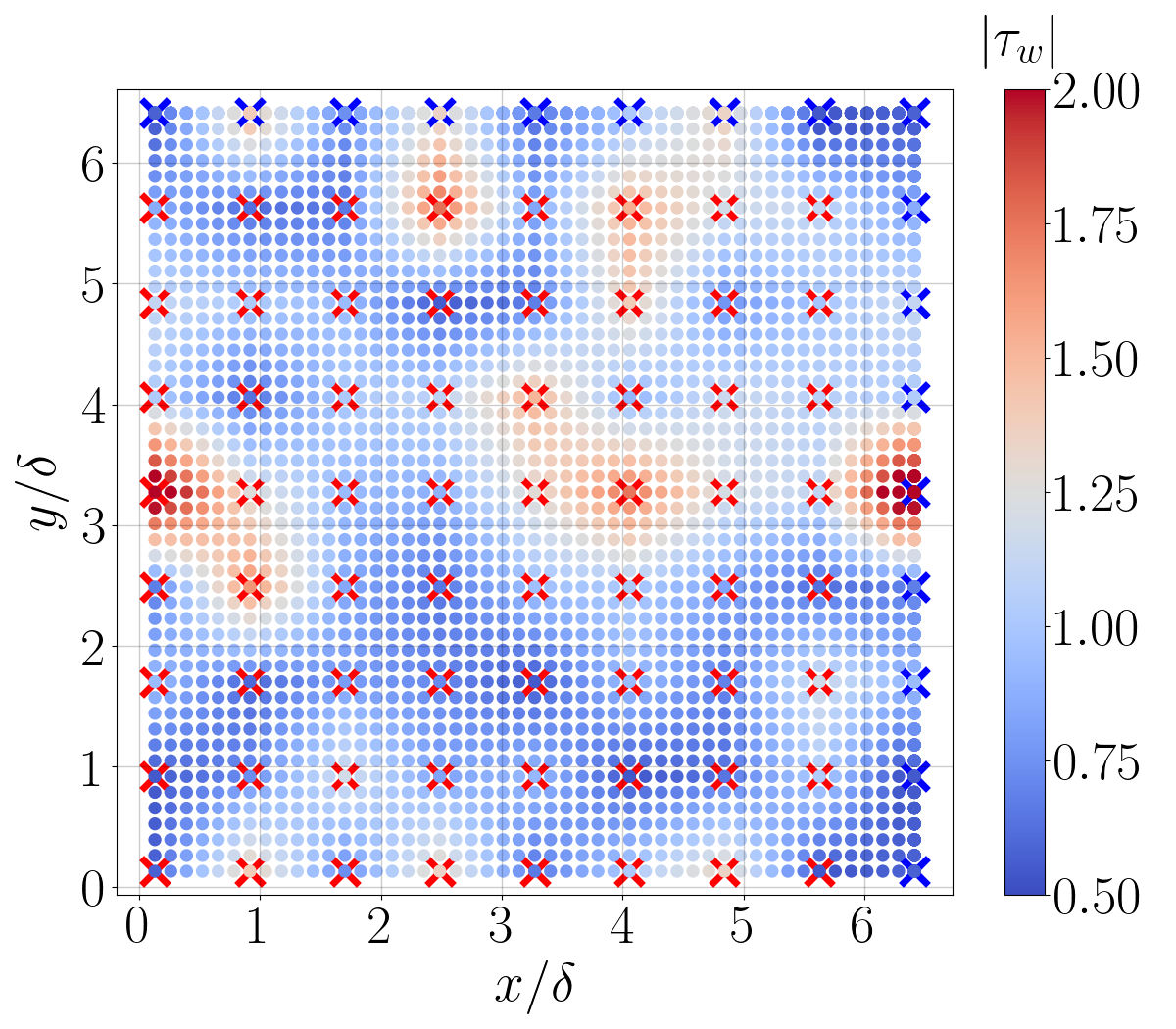}}\label{fig:Nag64}
\subfigure[$N_\textrm{agents}=128$.]{\includegraphics[width=0.325\textwidth]{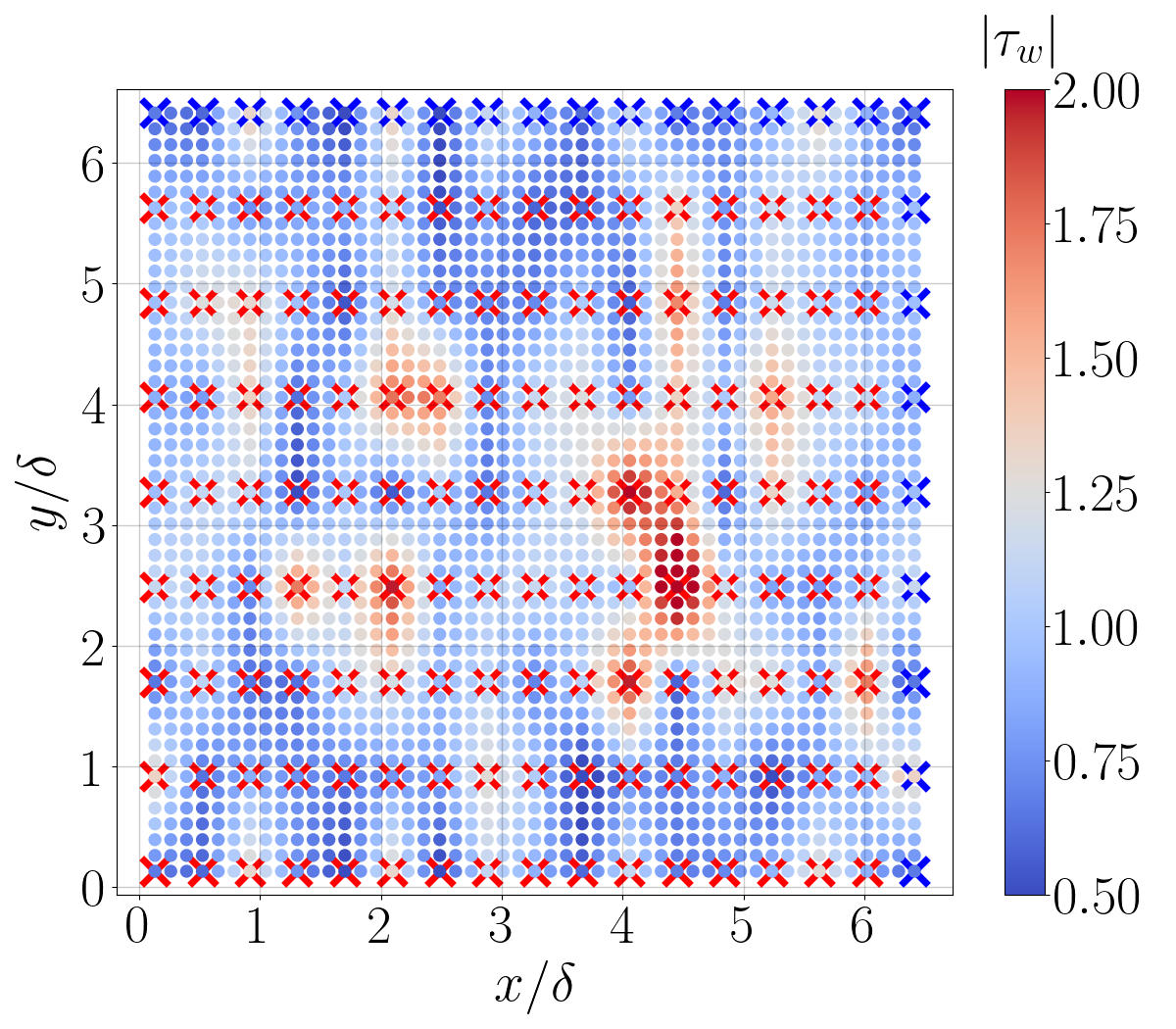}}\label{fig:Nag128}
\caption{A visualization of the instantaneous wall-shear stress when there are (a) 16, (b) $64$ and (c) $128$ agents.} \label{fig:Nag}
\end{figure}

\begin{figure}
\includegraphics[width=0.4\textwidth]{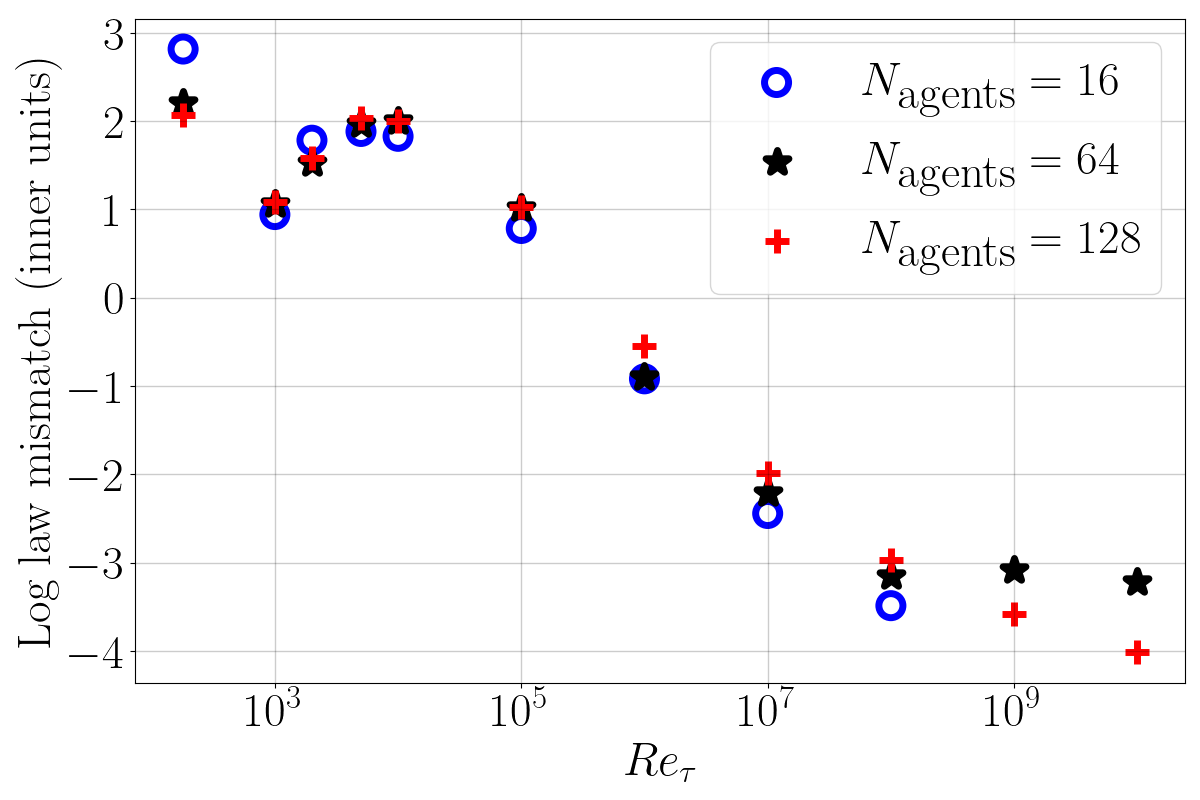}
\caption{\label{fig:mismatch} LLM as a function of the friction Reynolds number $Re_\tau$ for different number of agents.}
\end{figure}

Figure \ref{fig:Nag} displays wall-shear stress contours for a varying number of agents, $16$, $64$ and $128$ agents for BK22.
These agents are represented with red crosses in the figure.
Due to the periodic boundary conditions, we have extra 9, 17, and 25 agents at the boundaries.
These agents are represented with blue crosses.
The wall-shear stress at a point that does not have an agent is computed from a bi-linear interpolation. 
Apparently, reducing the number of agents results in less variations in the instantaneous wall-shear stress. 
Figure \ref{fig:mismatch} shows the LLM as a function of the friction Reynolds number for different numbers of agents. 
The number of agents has a negligible impact on the results. 
It is just that for $Re_\tau>10^9$, the simulation did not converge for 16 agents.
This is not an inadequacy of the model as 16 agents are too few. 

\begin{figure}[htbp]
\centering
\subfigure[EWM.]{\includegraphics[width=0.4\textwidth]{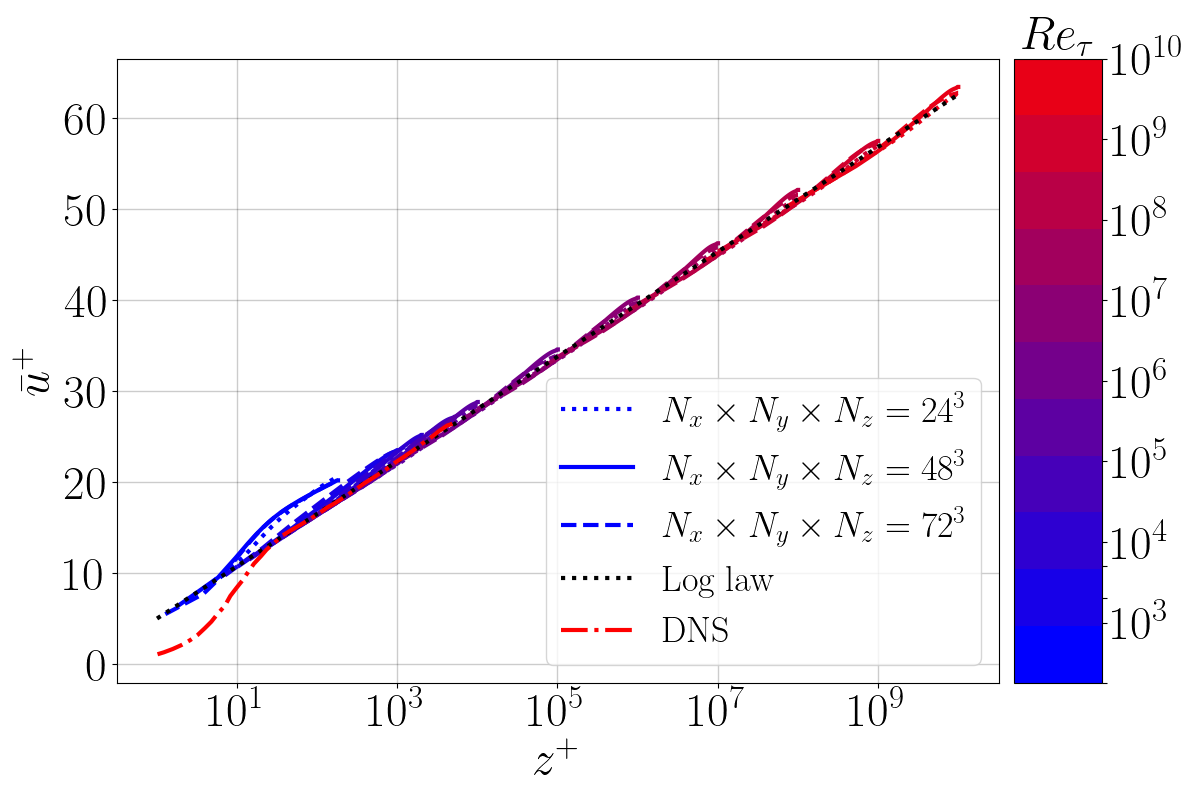}}
\subfigure[HYK19.]{\includegraphics[width=0.4\textwidth]{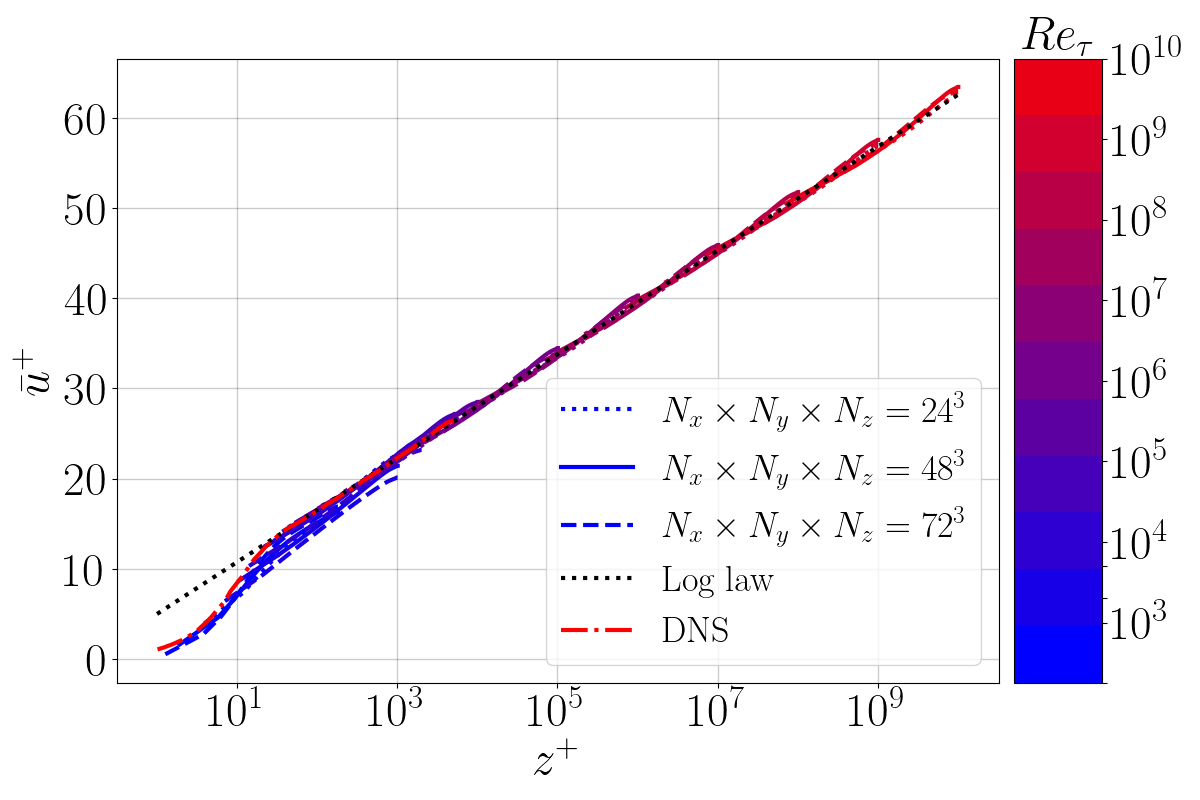}}
\subfigure[BK22.]{\includegraphics[width=0.4\textwidth]{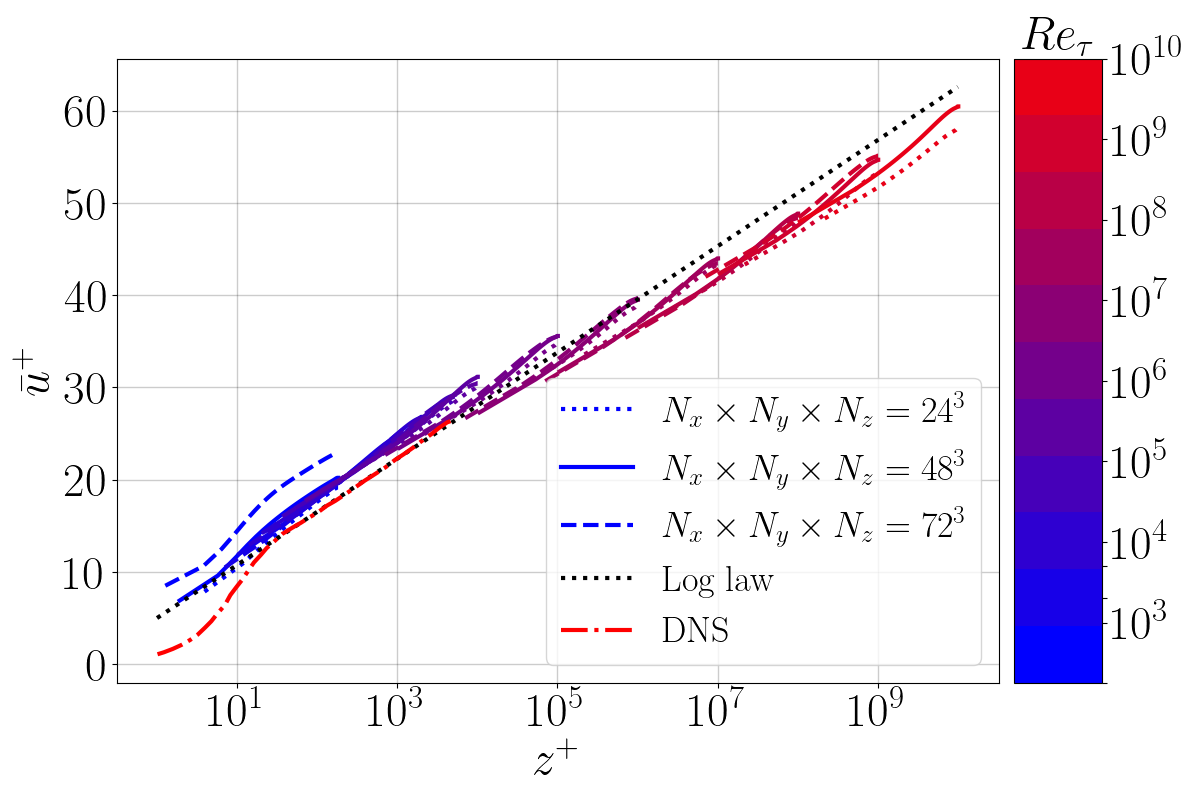}}
\caption{\label{fig:resolution}Mean velocity $\bar{u}^+$ as a function of the wall-normal coordinate $z^+$ plotted for $Re_\tau \in \left[180, 10^3, 2\times10^3, 5.2\times 10^3, 10^4, 10^5, 10^6, 10^7, 10^8, 10^9, 10^{10}\right]$. 
(a) EWM \cite{kawai2012wall}, (b) HYK19 \cite{huang2019wall}, (c) BK22 \cite{bae2022scientific}. 
Three different grid resolutions are considered, namely, $N_x \times N_y \times N_z =24^3$, $48^3$ and $72^3$. The BK22 simulation at $Re=10^{10}$ for $N_x \times N_y \times N_z =72^3$ did not reach convergence, likely because of too few agents (64) compared to the number of grid points. 
%the log law and with DNS results at $Re_\tau=5.2\times 10^3$ \cite{li2008public}.
} 
\end{figure}

Figure \ref{fig:resolution} shows the mean flow for grid resolutions $N_x \times N_y \times N_z = 24^3$, $48^3$, and $72^3$ for EWM, HYK19, and BK22. 
The grid resolution has negligible impact on the EWM and HYK19 results.
This is desired.
The BK22 results have some weak dependence on the grid resolution.
We will explain this in the next section.

\section{\label{sec:analyse} Analysis}

\black
The purpose of this section is to provide an analysis of the behavior of black-box MLWMs and to gain insights into their workings. 
This is not commonly done but we believe it will be instructive. 
The discussion will involve the extrapolation theorem \cite{bin2022progressive}, which governs the way in which a neural network extrapolates.
\black
The theorem reads:
for a non-trivial feed-forward neural network ``net'' that maps from $\mathscr{R}^1$ to $\mathscr{R}^1$, net$(\infty)=$ a finite constant if one employs the sigmoid transfer function for all neurons, and net$(x)\sim x$ (including net$(x)\sim 0\cdot x$) if one employs the rectified linear unit as the transfer function for all neurons.
\black
The theorem was initially developed for bias-free neural networks, but it is often applicable to general neural networks as well.

\black

\subsection{Supervised MLWM, HYK19}
\begin{figure}
\includegraphics[width=0.4\textwidth]{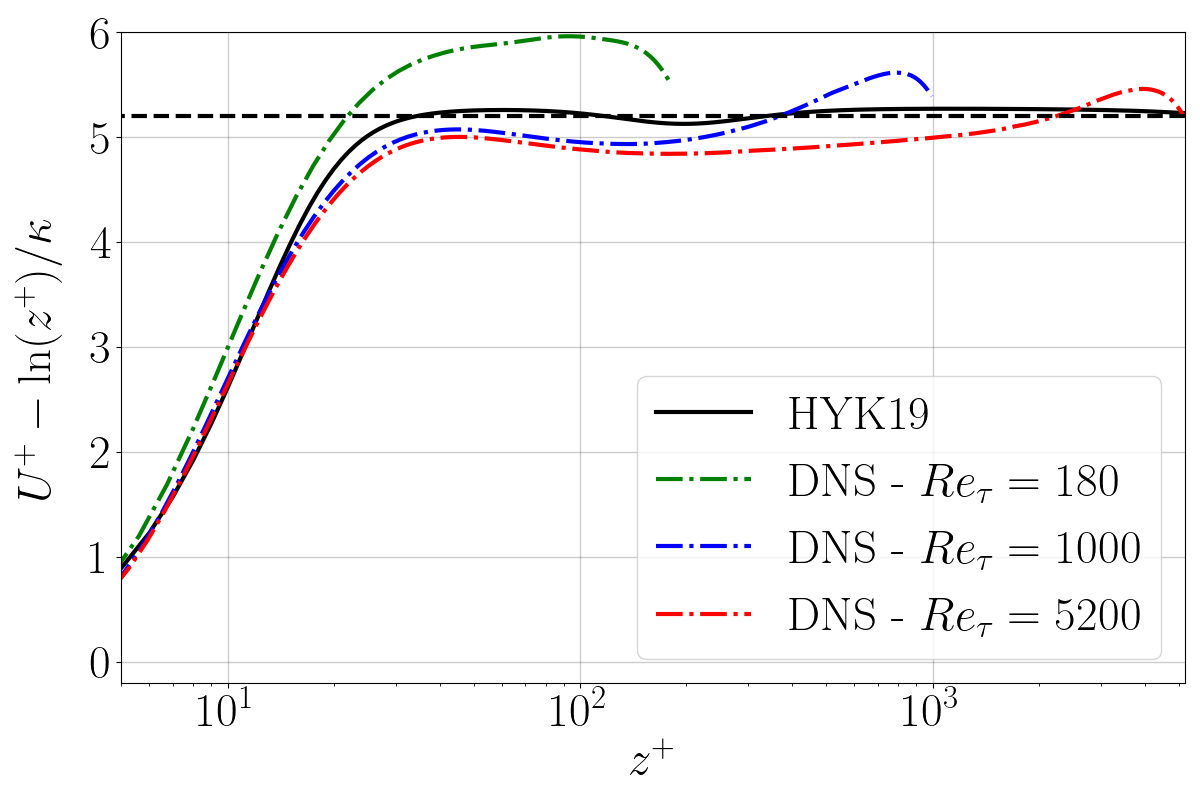}
\caption{\label{fig:interc} The network output $\left( U^+ - \ln(z^+)/\kappa\right)$ as a function of the input $z^+$ (in the absence of system rotation, the number of inputs reduces to 1). 
DNS data at $Re_\tau=180$, $Re_\tau=1000$ \cite{graham2016web} and $Re_\tau=5200$ \cite{lee2015direct} are included for comparison.
The black line is at $B\approx 5.2$, which is the typical value for the log-law intercept. 
\black The von Kármán constant is equal to $\kappa=0.4$. \black}
\end{figure}

\black
HYK19 was shown to give improved results compared to the EWM in spanwise rotating channels \cite{huang2019wall} and is shown in Section \ref{sec:result} to also preserve the law of the wall at seen $180<Re_\tau<1500$, and unseen $Re_\tau > 1500$, Reynolds numbers.
To gain insights into the behavior of the model, we compare its predictions with the law of the wall.
\black
The network is itself a model of the mean flow.
Figure \ref{fig:interc} shows the network's output $U^+-\ln(z^+)/\kappa$ as a function of its input $z^+$ in the absence of spanwise system rotation.
\black
We have observed two important characteristics of the model's behavior.
Firstly, the model is consistent with the law of the wall in the viscous layer up to $z^+=30$, indicating successful training of the network.
Secondly, the model asymptotes to a constant value at large $z^+$ values, preserving the law of the wall at high Reynolds numbers.
The exact value that the network asymptotes to is determined by the network's training, but the fact that it asymptotes to a constant is a result of the network's design, which uses a sigmoidal activation function. 
The extrapolation theorem states that a network with such a function will asymptote to a constant at infinity.

To summarize, the model presented in Ref. \cite{huang2019wall} is designed to preserve the law of the wall and offers improvements in and only in rotating channels.
\black
These improvements are not at the expense of its performance in other flows, where the model gives the same results as the EWM.

\subsection{Supervised MLWMs, ZHY21 and ZYZY22}

\black
In contrast to its superior performance in {\it a priori} tests, ZHY21 did not perform well in {\it a posteriori} tests.
\black
The model was originally trained on periodic hill flows, while we tested it on a different configuration, namely channel flow, and in a different solver. 
This difference in configuration and solver may explain why the model did not perform properly in our testing. 
On the other hand, ZYZY22 incorporated channel flow data generated from the law of the wall ($Re_\tau \in [10^3;10^9]$) in its training and was able to capture the law of the wall in WMLES but produces a von Kármán constant of $\kappa=0.32$.
\black
Although not shown in the previous section, ZYZY22's performance deteriorates at grid resolutions other than $N_x\times N_y\times N_z=32^3$, where the distances between the neighboring matching locations are not $0.03\delta$.

We analyze the consistency of WMs with the law of the wall to gain insights into their behavior in {\it a posteriori} tests.
\black
The law of the wall contains both the viscous layer and the logarithmic layer.
Here, we construct the law of the wall by stitching the wall layer in a $Re_\tau=5200$ channel \cite{lee2015direct} and the logarithmic law, $\bar{u}^+=\ln(z^+)/\kappa + B$, at $z^+=100$.
The flow rate, which is needed to compute the WMs's input, is obtained by integrating the law of the wall.
The two WMs take velocity and pressure information at three off-wall locations.
The law of the wall gives the velocity at any three off-wall locations.
The pressure terms are set to 0 per the constant stress layer assumption that underpins the law of the wall.

\begin{figure}
\centering
\includegraphics[width=0.4\textwidth]{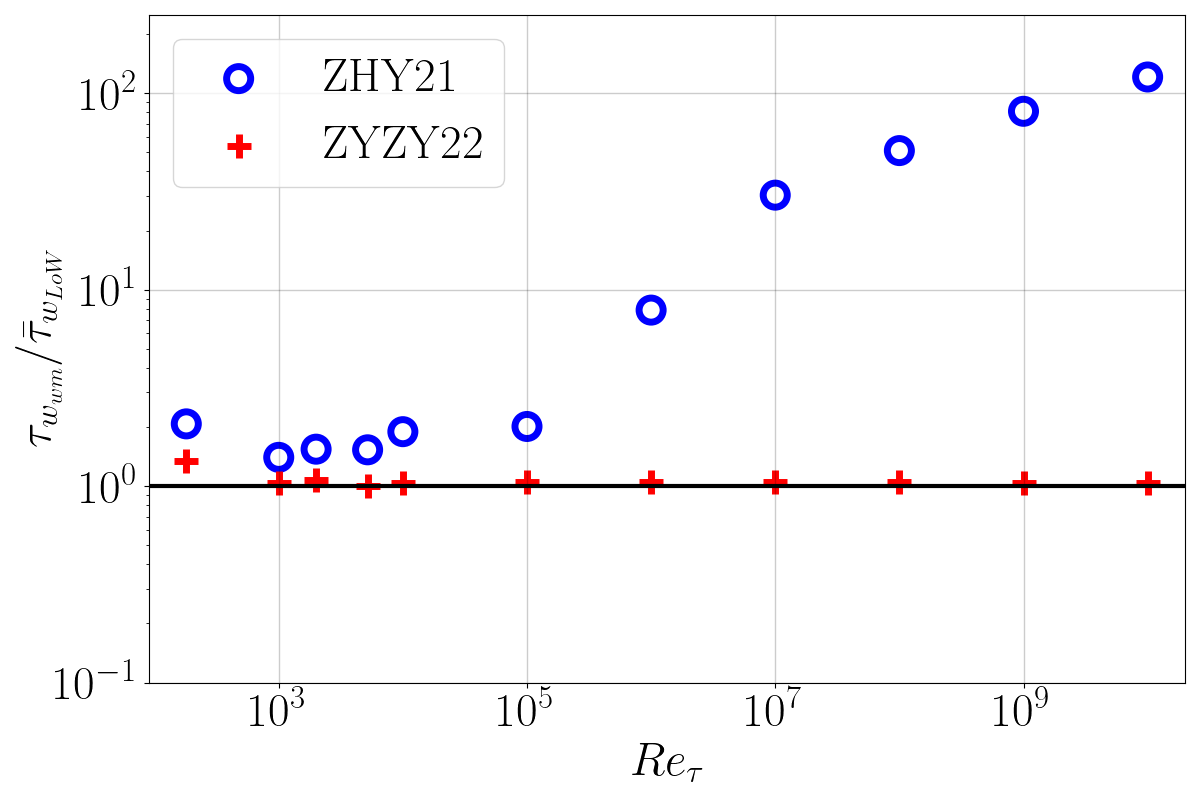}
\caption{\label{fig:v1v2} $\tau_{w_{wm}}/\tau_{w_{LoW}}$ as a function of the friction Reynolds number. Here, $\tau_{mw_{wm}}$ is the wall-shear stress given by ZHY21 or ZYZY22, $\tau_{w_{LoW}}$ is the wall-shear stress consistent with the law of the wall. } 
\end{figure}

For the first test, we take the velocity at $z_1=0.03\delta$, $z_2=0.06\delta$, and $z_3=0.09\delta$, and vary the Reynolds number.
Figure \ref{fig:v1v2} compares the predicted wall-shear stress and the truth.
ZHY21 overpredicts the wall-shear stress.
The error increases as the Reynolds number increases.
This explains its poor performance in {\it a posteriori} studies.
In contrast, ZYZY22 performs well at all Reynolds numbers except for $Re_\tau=180$.

It is a common practice to blame a lack of training data for the poor performance of a ML model and leave no comment on how much data is actually needed, which is not very helpful.
Here, ZHY21's poor performance is not entirely because of a lack of training data.
The rectified-linear unit (ReLU) activation function is also responsible.
Per the extrapolation theorem \cite{bin2022progressive}, as one of the network inputs, $\ln(h_{wm}/y^*)$, asymptotes to infinity at the infinite Reynolds number, the network output, $\tau_w/(\rho U_b^2)$, also asymptotes to infinity at the infinite Reynolds number.
This asymptotic behavior is erroneous because $\tau_w/(\rho U_b^2)$ should asymptote to 0 at the infinite Reynolds number.
The above explains the error in network ZHY21.
We add two caveats to this discussion.
\black
First, although the ReLU activation function is responsible for the behavior of the specific network ZHY21, it is not true that one cannot achieve the correct asymptotic behavior using ReLU. 
The extrapolation theorem does not preclude the asymptotic behavior and a proper designed NN can cancel features that asymptotes to infinity such as: $0\cdot x+{\rm constant}$ at $x=\infty$.
However, achieving this requires expertise in machine learning.
\black
Second, the extrapolation theorem was only known recently.
Zhou \emph{et al.} \cite{zhou2021wall} could not have known how to design a network to guarantee the right asymptotic behavior at the infinite Reynolds number.

\begin{figure}
\centering
\subfigure[$\tau_{xz_{wm}}/\bar{\tau}_{xz}$ with $h_{wm,1}=dz$ and $dh_{wm}=0.03\delta$]{\includegraphics[width=0.4\textwidth]{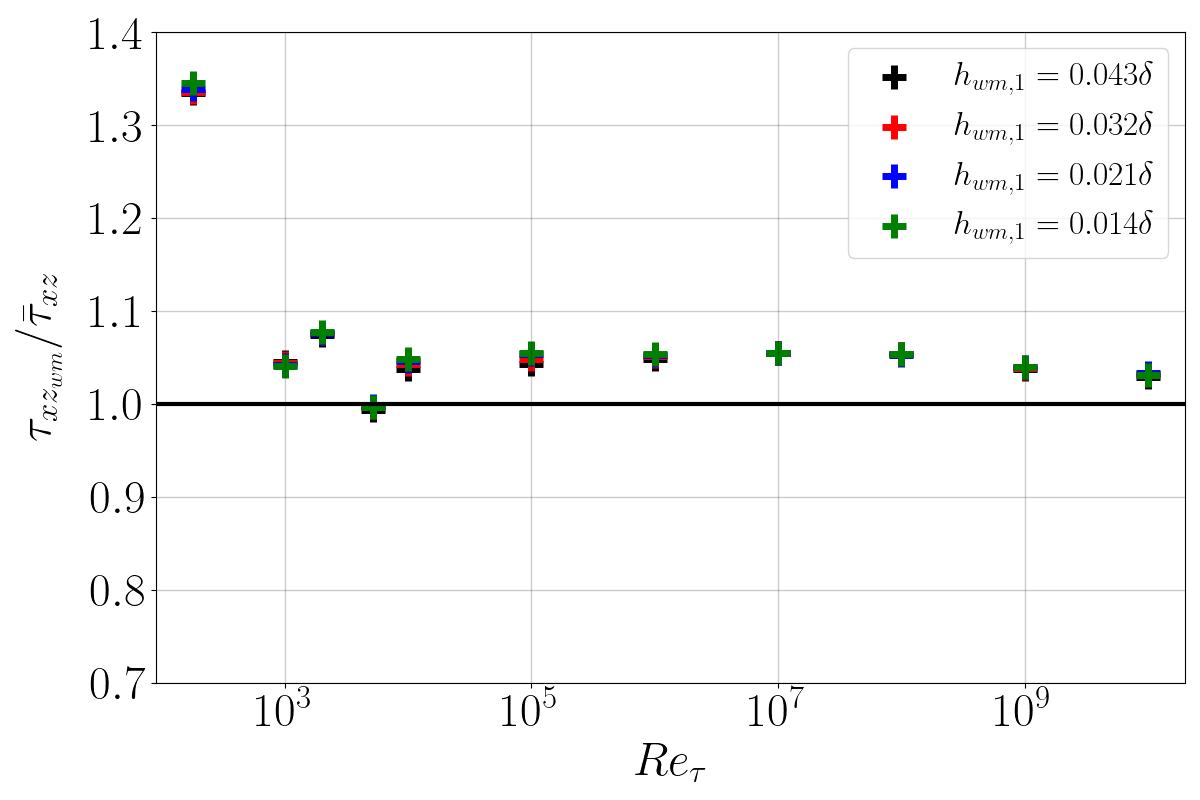}}
\subfigure[$\tau_{xz_{wm}}/\bar{\tau}_{xz}$ with $h_{wm,1}=dz$ and $dh_{wm}=dz$]{\includegraphics[width=0.4\textwidth]{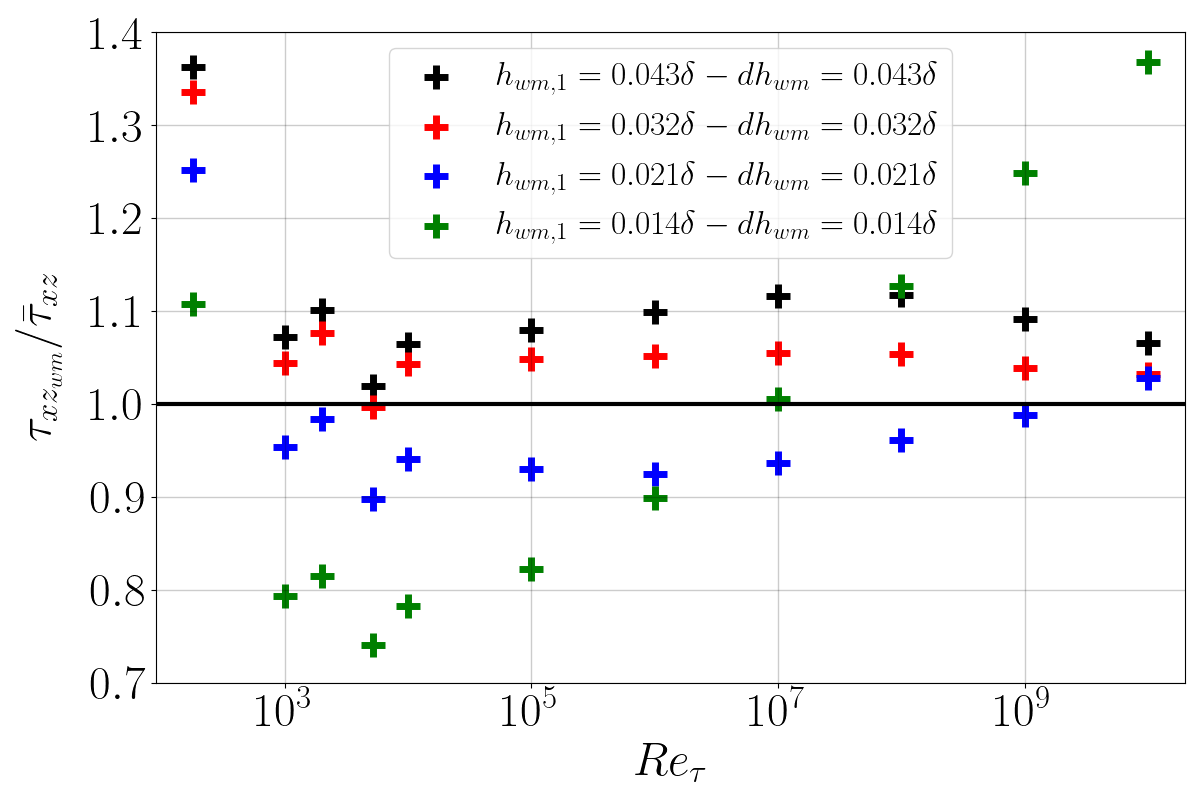}}
\caption{\label{fig:tau} $\tau_{xz_{wm}}/\bar{\tau}_{xz}$ as a function of the friction Reynolds number with (a) constant $dh_{wm}=0.03\delta$ and (b) $dh_{wm}=dz$. In (a) and (b), the location of the first off-wall point $h_{wm_1} = dz$.  } 
\end{figure}

For the second test, we vary $h_{wm,1}$, i.e., the distance between the wall and the first matching location and keep $dh_{wm}=0.03\delta$, i.e., the distances between two neighboring matching locations fixed at the training condition.
The results are shown in Figure \ref{fig:tau} (a).
We see that varying the distance between the wall and the first matching location does not incur any error. 
For the third test, we vary $h_{wm,1}$ and force $dh_{wm}=dz=L_z/(N_z-1)$, which deviates from the training conditions.
The results are shown in Figure \ref{fig:tau} (b), and errors show up.
These errors can be removed if one trains for $dh_{wm}$ at values other than $0.03\delta$ \cite{zhouArxiv}, which is outside the scope of this comparative study and left for future investigation.

\subsection{Reinforcement learning WM, BK22}
\label{sec:RL_analysis}

\begin{figure}
\centering
\subfigure[$Re_\tau=180$.]{\includegraphics[width=0.4\textwidth]{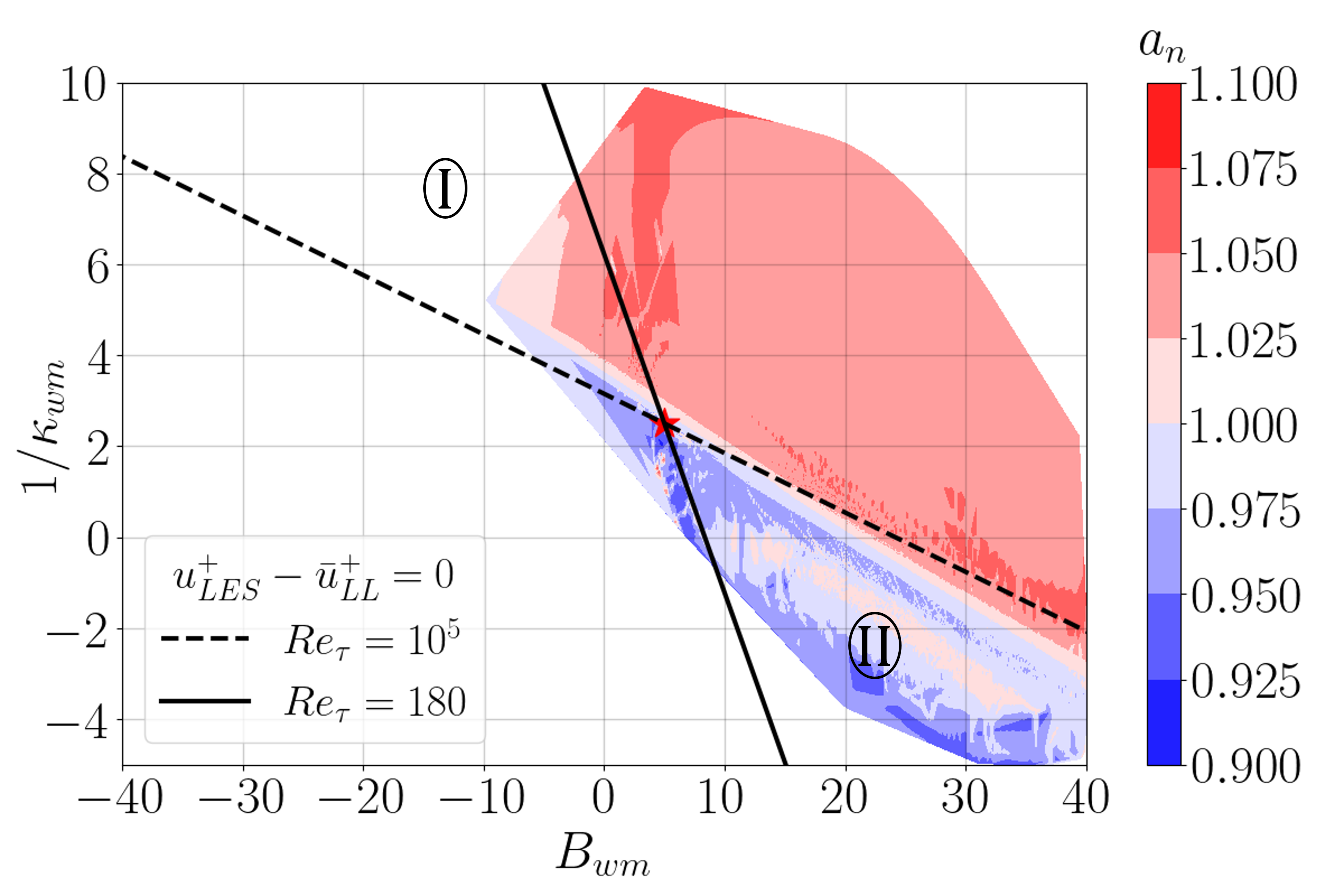}}
\subfigure[$Re_\tau=10^{3}$.]{\includegraphics[width=0.4\textwidth]{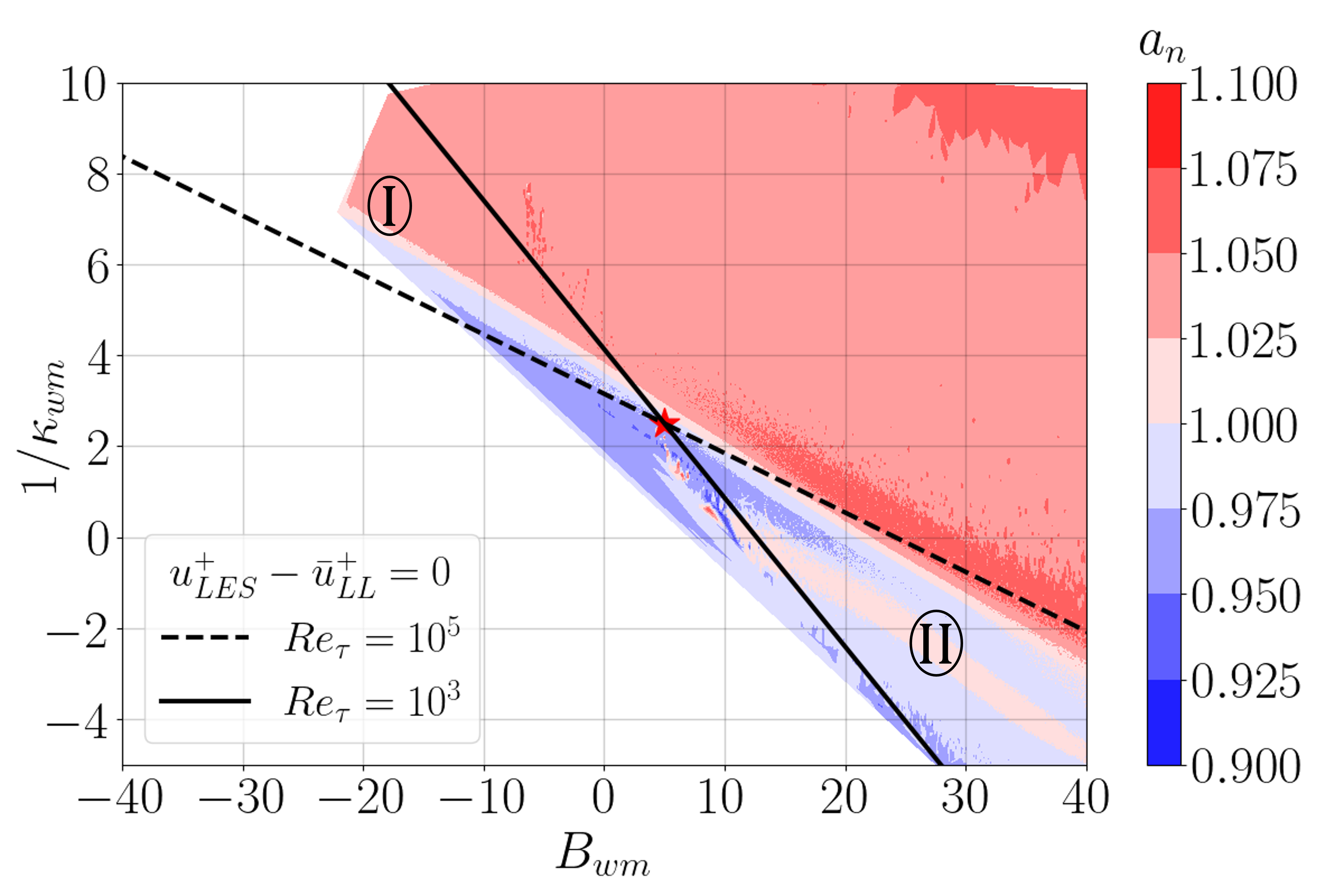}}
\subfigure[$Re_\tau=10^{5}$.]{\includegraphics[width=0.4\textwidth]{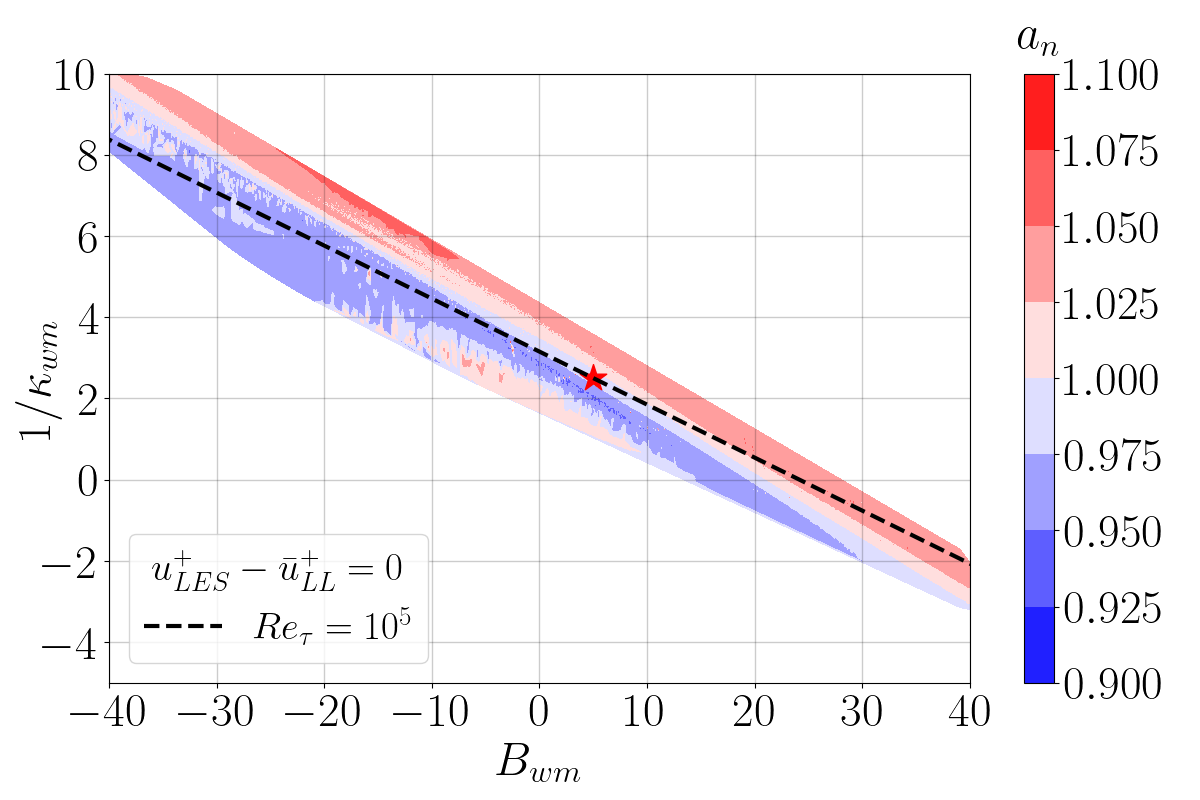}}
\subfigure[$Re_\tau=10^{10}$.]{\includegraphics[width=0.4\textwidth]{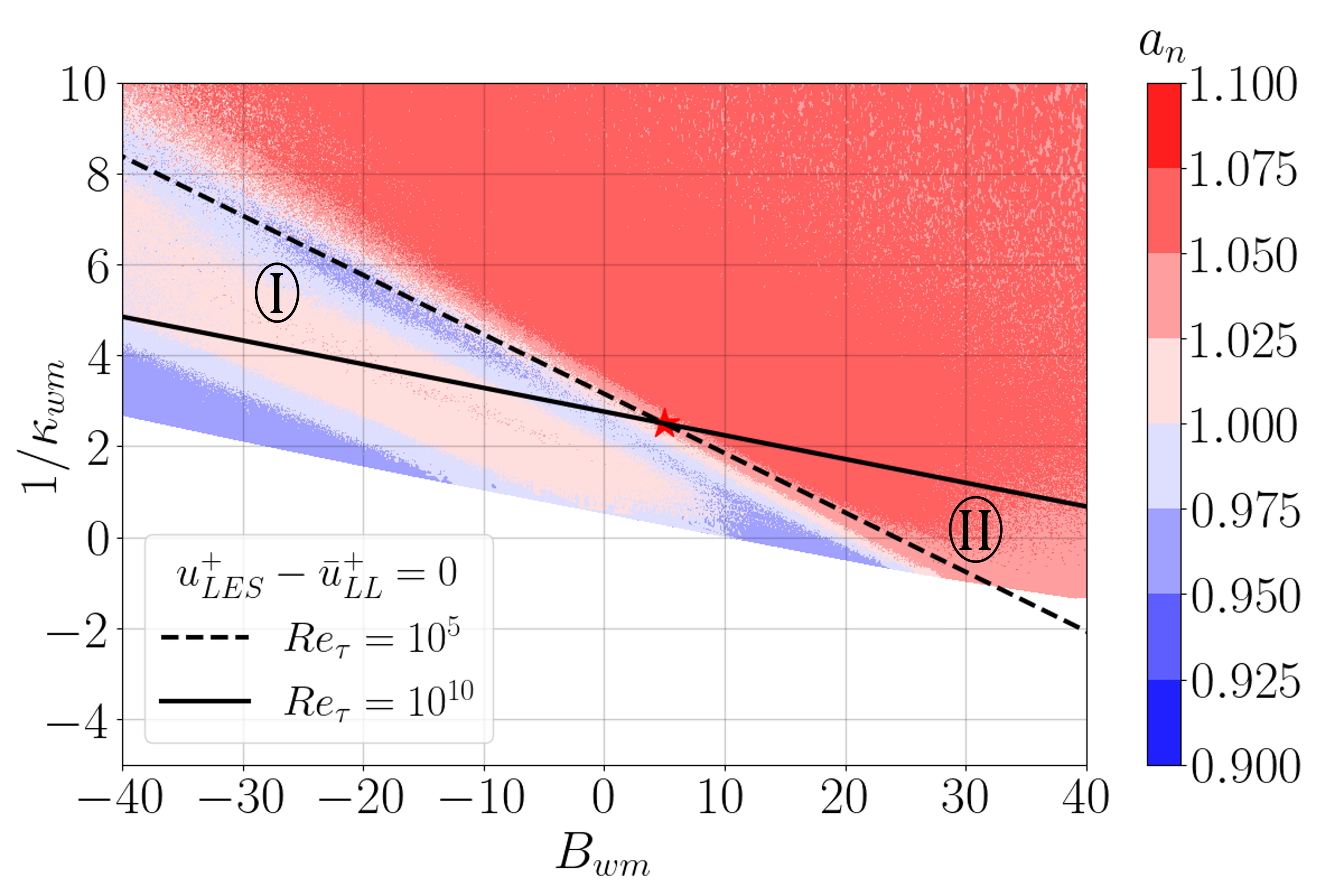}}
\caption{ \label{fig:analysis} State maps at (a) $Re_\tau=180$, (b) $Re_\tau=10^3$, (c) $Re_\tau=10^5$ and (d) $Re_\tau=10^{10}$. The contours show the actions. Data are collected from WMLES with the grid resolution $48^3$. } 
\end{figure}

BK22 captures the law of the wall with positive LLM at low Reynolds numbers and negative LLM at high Reynolds numbers.
Its behaviors are best understood by studying the state map or the action map.
Figure \ref{fig:analysis} shows the state maps for the $Re_\tau=180$, $10^3$, $10^5$ and $10^{10}$ channels, and we gather the actions taken by the RLWM during a statistically stationary period.

The state map contains a neutral line: \begin{equation}
u^+_{\rm LES}-\bar{u}_{\rm LL}^+ = 0.
\end{equation}
Here, $\bar{u}_{LL}^+$ is the velocity obtained from the log law with $\kappa=0.4$ and $B=5$.
Both velocities are evaluated at the matching location $h_{wm}$. 
For states located above this neutral line, the velocity is larger than the log-law value. 
Given these states, the RLWM should ideally generate an action $a_n>1$. 
By doing so, the wall-shear stress would increase, which is anticipated to result in a drop in the local velocity, thereby bringing down the velocity to the log-law value. 
On the contrary, given states below the neutral line, the RLWM should ideally generate an action $a_n<1$ to bring up the velocity to the log-law value.
In all, there may be four scenarios: 
\begin{itemize}
    \item A = \{$u^+_{wm}>\bar{u}_{\rm LL}^+$ \& $a_n>1$\};
    \item B = \{$u^+_{wm}>\bar{u}_{\rm LL}^+$ \& $a_n<1$\};
    \item C = \{$u^+_{wm}<\bar{u}_{\rm LL}^+$ \& $a_n>1$\};
    \item D = \{$u^+_{wm}<\bar{u}_{\rm LL}^+$ \& $a_n<1$\};
\end{itemize}    
among which A and D are desired.

\begin{figure}
\centering
\subfigure[]{\includegraphics[width=0.4\textwidth]{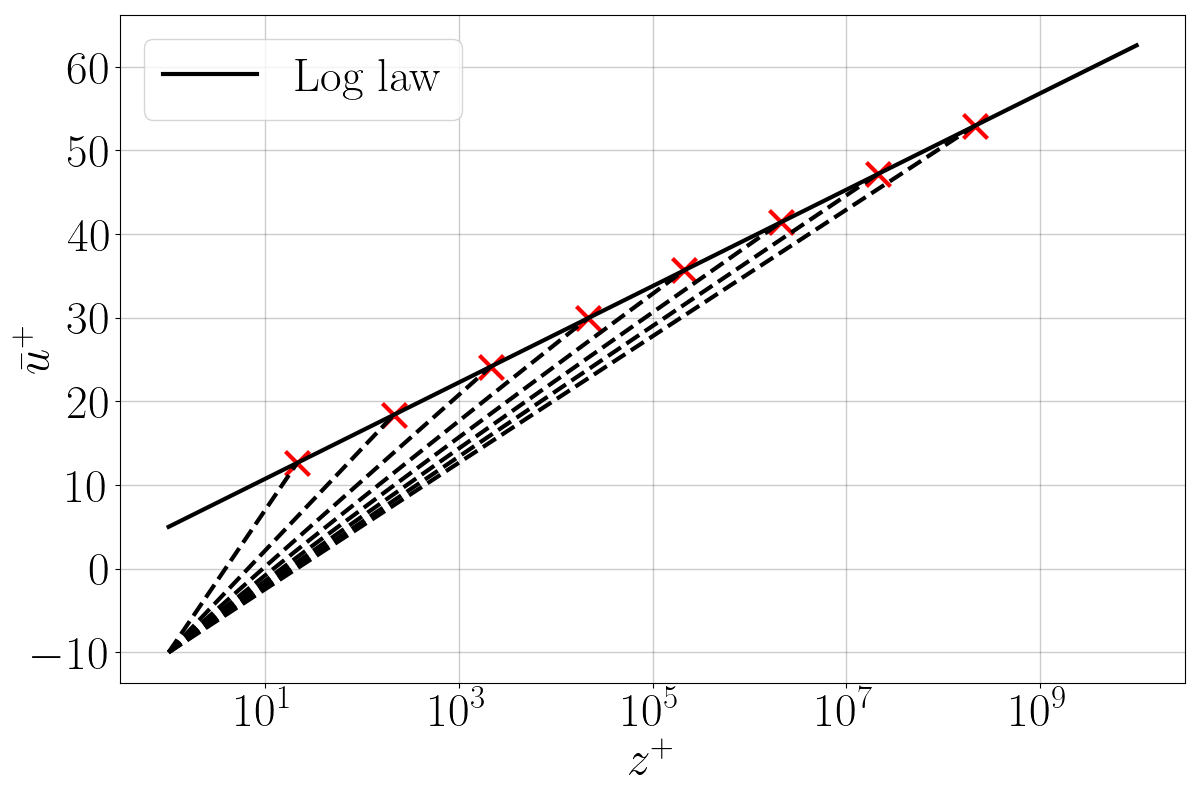}}
\subfigure[]{\includegraphics[width=0.4\textwidth]{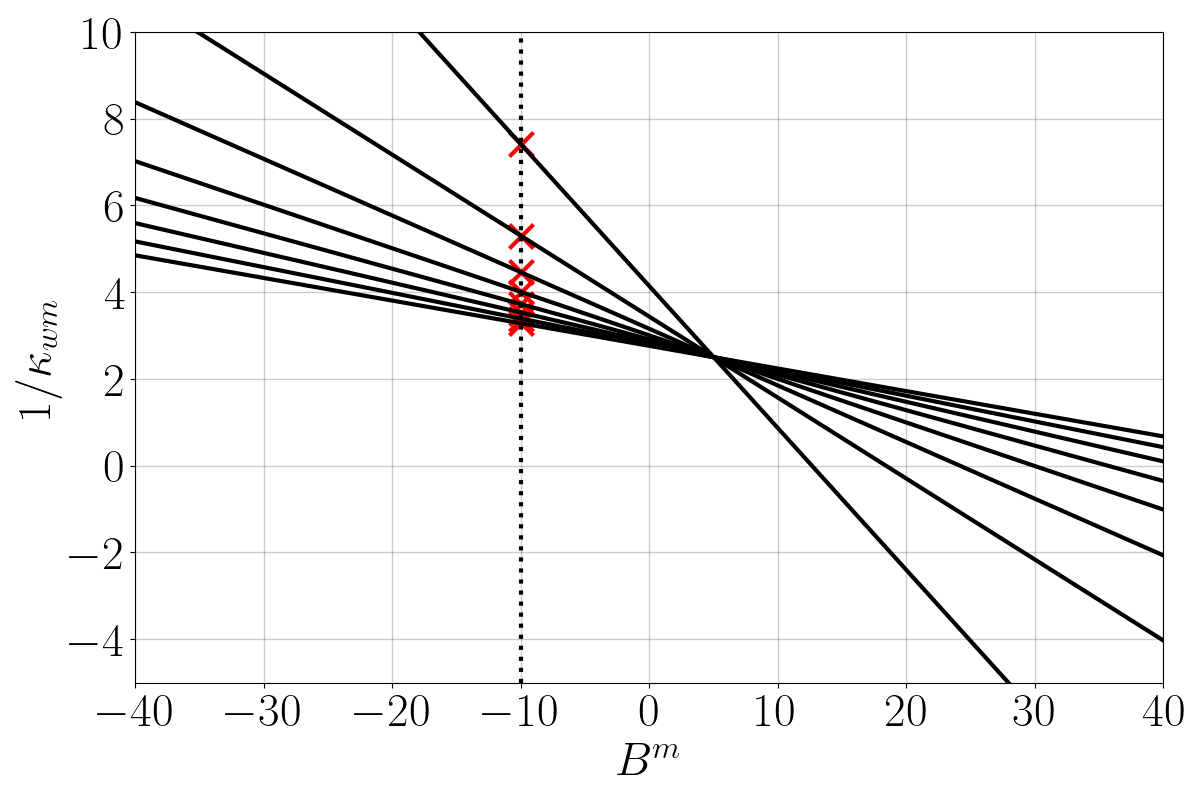}}
\caption{\label{fig:expl} 
(a) Profiles corresponding to $u^+_{\rm LES}-\bar{u}_{\rm LL}=0$.
(b) Corresponding neutral lines in the state map.
} 
\end{figure}

We can re-write the neutral line as follows:
\begin{equation}
    \frac{1}{\kappa_{wm}}\ln(h_{wm}^+)+B_{wm} - u^+_{\rm LL}=0,
\end{equation}
alternatively,
\begin{equation}
    \frac{1}{\kappa_m}=\left(-\frac{1}{\ln(dz/\delta)+\ln(Re_\tau)}\right)B_{wm}+\frac{u^+_{\rm  LL}}{\ln(h_{wm}^+/Re_\tau)+\ln(Re_\tau)},
    \label{eq:neutral}
\end{equation}
where $-1/(\ln(dz/\delta)+\ln(Re_\tau))$ is the slope of the neutral line and ${u^+_{\rm  LL}}/({\ln(h_{wm}^+/Re_\tau)+\ln(Re_\tau}))$ is the intercept.
It follows from Eq. \eqref{eq:neutral} that the slope of the neutral line decreases as the Reynolds number increases and increases as the grid resolution increases, the latter of which explains our observation in Figure \ref{fig:resolution}. 
This is more clear from Figure \ref{fig:expl}.
Figure \ref{fig:expl} (a) shows the profiles that correspond to $u^+_{\rm LES}-\bar{u}^+_{\rm LL}=0$ and Figure \ref{fig:expl} (b) shows the corresponding neutral lines in the state map.
For a given grid resolution $dz/\delta$, $h_{wm}^+$ grows as $Re_\tau$ increases. 
Hence, given $B_{wm}<B$, the log-law slope decreases as the Reynolds number increases in order to satisfy the condition $u^+_{wm}=\bar{u}^+_{LL}$, which corresponds to an increasing $\kappa_{wm}$. 
Also, as is clear from Figure \ref{fig:expl}, the log-law slope depends less sensitively on the Reynolds number as the Reynolds number increases.

The neutral line should ideally separate the $a_n>1$ actions and the $a_n<1$ actions, which is the case in Figure \ref{fig:analysis} (c).
\black
However, this is not always the case.
\black
Figure \ref{fig:analysis} (a, b, d) shows the state maps at $Re_\tau=180$, $1000$, and $10^{10}$. 
The neutral lines in these three plots are at different positions than that in Figure \ref{fig:analysis} (c) but the line that separates the $a_n>1$ and the $a_n<1$ actions remains about the same.
Consequently, the RLWM generates $a_n<1$ and $a_n>1$ actions in regions I and II, respectively, which brings the velocities that are above and below the log-law values further away from the log-law values, resulting in a negative log layer mismatch at $Re_\tau=10^{10}$ and a positive log layer mismatch at $Re_\tau=180$ and 1000.
\black 
Considering our resolution, the matching location ($h_{wm}^+$) for the $Re_\tau=10^5$ case is near the matching location used for the original training, which included channel flows at $Re_\tau=2000$, $4200$, and $8000$. 
This is why the performance of the model is best near this Reynolds number range, but we observed a mismatch outside this range.
\black

We may take more quantitative measures like the precision and the recall, which are defined for action $a_n>1$ as: 
\begin{equation}
\begin{split}
    \text{precision}(a_n>1) & =\frac{\text{true positive}}{\text{true positive}+\text{\bf false positive}}= \frac{\text{instances of A}}{\text{instances of A}+\text{instances of C}},\\
    \text{recall}(a_n>1) & =\frac{\text{true positive}}{\text{true positive}+\text{\bf false negative}}= \frac{\text{instances of A}}{\text{instances of A}+\text{instances of B}},\\
    %\text{F1 score} &= \sqrt{\text{precision}\times \text{recall}}.
\end{split}
\end{equation}
Similarly, one can define the precision and recall for action $a_n<1$.
We can also define accuracy: 
\begin{equation}
    \text{Accuracy}=\frac{\text{instances of A and D}}{\text{all instances}}.
\end{equation}
These measures are collected in Table \ref{tab:Recall} for all Reynolds numbers. 
The accuracy is about 0.50 at all Reynolds numbers. 
Hence, the RLWM generates a wall-shear stress that brings the local velocity closer to the log-law velocity for 50\% of the instances.
We note that 50\% accuracy is not necessarily bad because turbulence is itself stochastic.
The recalls for both the $a_n<1$ actions and the $a_n>1$ actions are about 0.5.
Hence, for about half of the instances that the velocity is above/below the log-law value, the RLWM generates a wall-shear stress that brings the velocity closer to the log-law value.
\black
The LLM is not strongly correlated with either the accuracy or the recall, but it is correlated with the precision. 
\black
When the precision of the $a_n>1$ actions is low, more instances of C (as compared to the instances of A) happen.
When these events happen, $\tau_w$ is increased in spite of $u^+_{wm}<\bar{u}^+_{LL}$, producing negative mismatch at $Re_\tau\geq 10^7$. 
On the other hand, when the precision of the $a_n<1$ actions is low, more instances of B occur, where $\tau_w$ is decreased in spite of $u^+_{wm}>\bar{u}^+_{LL}$, leading to positive mismatches at $Re_\tau$ between 2000 and $10^4$.
The precisions for both the $a_n>1$ and the $a_n<1$ actions are low at $Re_\tau=180$ and $10^3$ and the effects cancel, leading to a small LLM.

\begin{table*}%[htbp]
\begin{ruledtabular}
\caption{\label{tab:Recall} Precision, recall and accuracy with the corresponding mismatches.}
\begin{tabular}{ccccccc}
 & \multicolumn{2}{c}{Precision} & \multicolumn{2}{c}{Recall} & & \\
$Re_\tau$ & $a_n>1$ & $a_n<1$ & $a_n>1$ & $a_n<1$ & Accuracy & \makecell{Mismatch\\(inner units)} \\
\hline
$180$ & 0.61 & 0.38 & 0.48 & 0.51 & 0.49 & +2.2 \\
\hline
$10^3$ & 0.61 & 0.40 & 0.49 & 0.52 & 0.50 & +1.1 \\
\hline
$2\times10^3$ & 0.68 & 0.34 & 0.49 & 0.53 & 0.51 & +1.5 \\
\hline
$5.2\times10^3$ & 0.80 & 0.25 & 0.50 & 0.57 & 0.51 & +2.0 \\
\hline
$10^4$ & 0.92 & 0.10 & 0.49 & 0.59 & 0.50 & +2.0\\
\hline
$10^5$ & 0.66 & 0.40 & 0.51 & 0.55 & 0.53 & +1.0\\
\hline
$10^6$ & 0.58 & 0.47 & 0.52 & 0.53 & 0.52 &-0.9 \\
\hline
$10^7$ & 0.55 & 0.48 & 0.51 & 0.52 & 0.51 & -2.2\\
\hline
$10^8$ & 0.56 & 0.47 & 0.52 & 0.52 & 0.52 & -3.2\\
\hline
$10^9$ & 0.57 & 0.44 & 0.49 & 0.52 & 0.50 & -3.1 \\
\hline
$10^{10}$ & 0.57 & 0.42 & 0.45 & 0.53 & 0.49&-3.2 \\
\end{tabular}
\end{ruledtabular}
\end{table*}

\subsection{Computational cost}
\black
The cost of a CFD model is important as well. 
A comprehensive analysis of the models' cost would necessitate a separate study examining performance across different platforms and codes, which is outside the scope of this work. 
Here, we follow the previous authors and provide a rough estimate, see, e.g., 
Refs. \cite{park2014improved,park2016numerical}. 
Specifically, we compare the costs of the three MLWMs to that of the ODE-based equilibrium wall model.
We will show that the three ML models are all cheaper than the ODE-based equilibrium wall model.
Since the cost of the ODE-based equilibrium wall model is not an issue, the costs of the three MLWMs should not be an issue either.
\black
It takes 5 operations to evaluate the algebraic EWM Eq. \eqref{eq:EWM} one time: a multiplication, two divisions, a logarithmic function evaluation, and a square.  
The number of operations needed to propagate information through a feed-forward neural network from the input layer to the output layer is given by:
\begin{equation}
    N=\sum_{i=2}^{L} (n_{i-1}+2)n_i,
    \label{eq:cost}
\end{equation}
where $L$ is the number of layers including the input and the output layer, $n_i$ is the number of neurons in the i\textsuperscript{th} layer excluding the bias unit, $N$ is the total number of operations, and +2 is for addition of the bias and the evaluation of the activation function.
This is a rough estimate: multiplication, addition, and the evaluation of the activation function are all considered as one operation.
It follows from Eq. \eqref{eq:cost} that evaluating HYK19, ZYZY22, and BK22 one time takes 56, 1609, and 17412 operations (assuming one agent per grid point).
\black
The above analysis shows that BK22 is the most costly among the three MLWMs---if it is applied at every location.
Next, we compare BK22 to the ODE-based equilibrium wall model.
The ODE-based equilibrium wall model adds about 10\% to 30\% overhead to the LES \cite{park2014improved,park2016numerical}, which is typically not a concern.
We compare the cost of the ODE-based wall model and BK22 on the PSU-ACI HPC cluster by independently evaluating these two models.
%Intel Xeon E5-2680 processors
The test reveals that the cost ratio between the ODE-based equilibrium wall model and BK22 is approximately four, indicating that BK22 is more cost-effective than the ODE-based equilibrium wall model.

Note that this analysis only considers the execution cost of the model and does not include the training cost. 
Similar to other readily available turbulence models, these MLWMs are designed to be used as they are, without the need for re-training. 
Thus, from the perspective of a WMLES user, training costs are not apparent.
That being said, it is important to acknowledge that for supervised WMs like HYK19 and ZYZY22, training demands the generation of a high-fidelity database in addition to the step of adjusting weights and biases. 
The associated cost is tied to the size of the training database and the neural network.
The training of RL models is generally quite expensive, involving a significant number of gradient steps (around $10^7$ for BK22). 
However, it does not require the creation of a high-fidelity database; instead, it requires a significant amount of WMLES for trial and error processes.
This comparative study targets WMLES users rather than developers, with the goal to facilitate the selection of an appropriate existing model. 
Given that these users do not perceive the training costs, we refrain from discussing them in further detail.
\black

\begin{comment}
\black
However, this is a rough estimate as the network in HYK19 needs to be evaluated multiple times to obtain a wall-shear stress, and using one agent at every wall location for BK22 WM is excessive.

Note that even if the MLWM generates a thousand times more operations than the EWM, it is not a problem if the WM computation is only a small part of the LES. 
WMLES using MLWMs are more efficient than WRLES, especially in industrial configurations with complex geometry and a larger number of grid points. 
Moreover, WRLES costs will increase with Reynolds number. 
These references \cite{park2014improved,park2016numerical} confirm that WMLES is significantly more cost-effective than WRLES, even when using wall models that are considerably more expensive than EWM.

Finally, it is worth noting that for the BK22 WM, library coupling is only required for training the RLWM and not necessarily for testing it.
The RLWM can be represented as one or more neural networks with weights and biases. 
A more computationally efficient approach to using these types of models would be to export the network as a matrix of weights and biases and perform matrix computations with GPUs for example. 
The authors believe that it is essential to first evaluate the benefit of this type of WMs and determine if it can outperform EWM before attempting to optimize its implementation and make it cost-effective.

\black
\end{comment}

\section{\label{sec:conclusion} Concluding remarks}

This study surveys the available MLWMs: HYK19, the supervised MLWM in Ref. \cite{huang2019wall}, ZYZY22, the supervised MLWM in Refs. \cite{zhou2021wall,zhouArxiv}, and BK22, the RLWM in Ref. \cite{bae2022scientific}. 
The implementation of the three WMs are made available in the open-source code LESGO so that anyone can pick up these ML models and use them for predictive modeling.

This study emphasizes the canonical channel flow.
We follow Ref. \cite{rumsey2022search} and argue that a ML model must preserve the known empiricism.
We vary the friction Reynolds number form $Re_\tau=180$ to $10^{10}$, and the MLWMs are compared with the baseline EWM and the law of the wall.
Among the three MLWMs, HYK19 gives accurate mean flow predictions at all Reynolds numbers, ZYZY22 captures the law of the wall, but predicts a smaller von Kármán constant, BK22 captures the law of the wall with positive and negative LLMs at low and high Reynolds numbers, respectively.
\black
The costs of the three MLWMs are all lower than the ODE-based equilibrium wall model and therefore are not a concern.
\black

In addition to documenting the results, we also attempt to explain why the MLWMs behave the way they behave.
This is rarely done but is important to modeling. 
{\it A priori} analysis shows that HYK19 and ZYZY22 both preserve the law of the wall.
This explains HYK19's good performance meanwhile points to LES (LES solver, SGS model) for the errors (the small von Kármán constant) in ZYZY22.
The analysis also shows that the LLMs in BK22's results are due to the mismatch between the neutral line and the line that separates the $a_n>1$ and $a_n<1$ actions.
When the two lines match, BK22 gives good results.
LLM arises when the two lines do not match. 
We show that when the neutral line is above the line that separates the $a_n>1$ and $a_n<1$ actions, positive LLM arises, and vice versa.

Rumsey \emph{et al.} \cite{rumsey2022search} argued that the improvements offer by a ML model in some flow should not be at the expense of other flows, requiring that a ML model must not degrade the accuracy of the baseline model.
\black We share this viewpoint and support a progressive learning approach where MLWMs are developed in a way similar to empirical WMs, by gradually increasing complexity and verifying that at each step of the process, the fundamental physical laws are accurately recovered. 
However, we should note that applying this viewpoint to all MLWMs, particularly new MLWMs, is not always fair.
Many papers aim to validate a new, novel methodology rather than to develop a CFD product.
The novelty of BK22 is that it requires no high-fidelity DNS data.
The novelty of HYK19 is that it accurately predicts rotating channel flows.
The novelty of ZYZY22 is that it takes information at multiple locations, to predict flow separation in periodic hills flows.
\black
We conducted this survey to evaluate the capability of MLWMs to recover EWM results on basic channel flow configurations beyond their training range.
However, the results are somewhat mixed. 
%We acknowledge that these methods might have the potential to outperform empirical WMs.
%Recent advances in fields such as games and natural language processing have demonstrated tremendous potential for ML methods. 
%For instance, RL has shown the ability to handle very large dimensional problems, such as in the game of Go, with up to $10^{170}$ states \cite{maddison2014move}.
%In turbulence modeling, which is also a high dimensional problem due to the wide range of scales that increases with the Reynolds number, ML tools are still in the early stages of development.
Nonetheless, there are still many ML methods that have not been explored yet, and we believe that with further research and development, MLWMs will yield better results.
Such an approach could prove particularly valuable in predicting complex flow problems, including separated flows, where significant progress could be made \cite{hansen2023pod}.
To encourage the acceptance and utilization of MLWMs, we follow \cite{rumsey2022search} and suggest placing emphasis on two key aspects: verifying their extrapolation capabilities, specifically their ability to recover fundamental physical laws and improving their interpretability. 
By doing so, we can better understand their potential and ensure their practical use in the field. 
%Regarding the cost of these methods, we believe that the priority should be to demonstrate the benefits and the ability to recover baseline laws before focusing on optimizing their computational efficiency.

\black

A new turbulence model must survive many independent comparative studies before it would be picked up and used for predictive modeling.
The field of MLWM is still young, and there will be many more comparative studies in the future.
This study is limited to channel flow, and we examined three available models.
The hope is that first, more easy-to-implement MLWMs would become publicly available; second, simple channel flow becomes a sanity check; third, there would be a consensus on the inputs needed to accurately predict the wall-shear stress; and last, MLWMs can help address the challenge in more complex flows such as separated flows.

\begin{acknowledgments}
This research is supported by the Independent Research Fund Denmark (DFF) under the Grant No. 1051-00015B.
Yang acknowledges US Office of Naval Research under contract N000142012315, with Dr. Peter Chang as Technical Monitor. 
The authors thank H. Jane Bae, Di Zhou, Xiaolei Yang, Zhideng Zhou, Kevin P. Griffin, and Michael P. Whitmore for their assistance and insightful remarks on the paper.

\end{acknowledgments}

\appendix

\black

\section{Close comparison with DNS}
\label{app:DNS}

\begin{figure}
\centering
\subfigure[EWM.]{\includegraphics[width=0.4\textwidth]{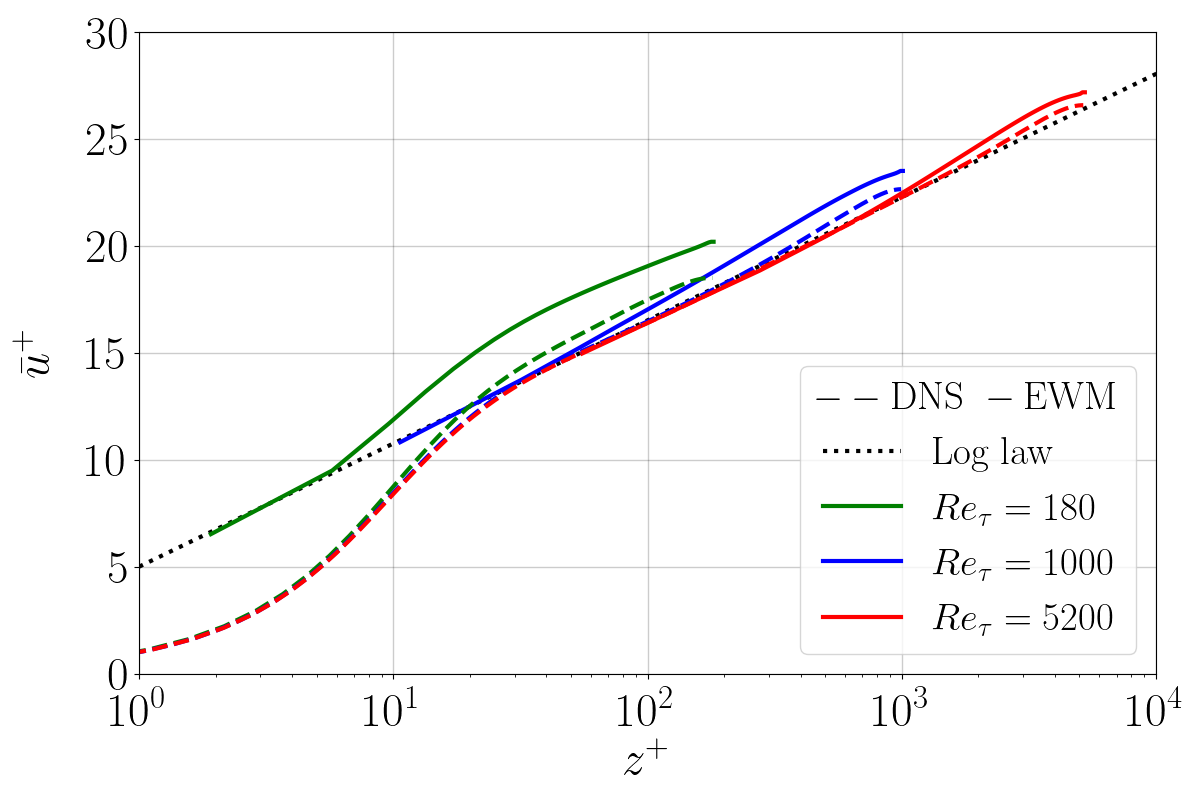}}
\subfigure[HYK19.]{\includegraphics[width=0.4\textwidth]{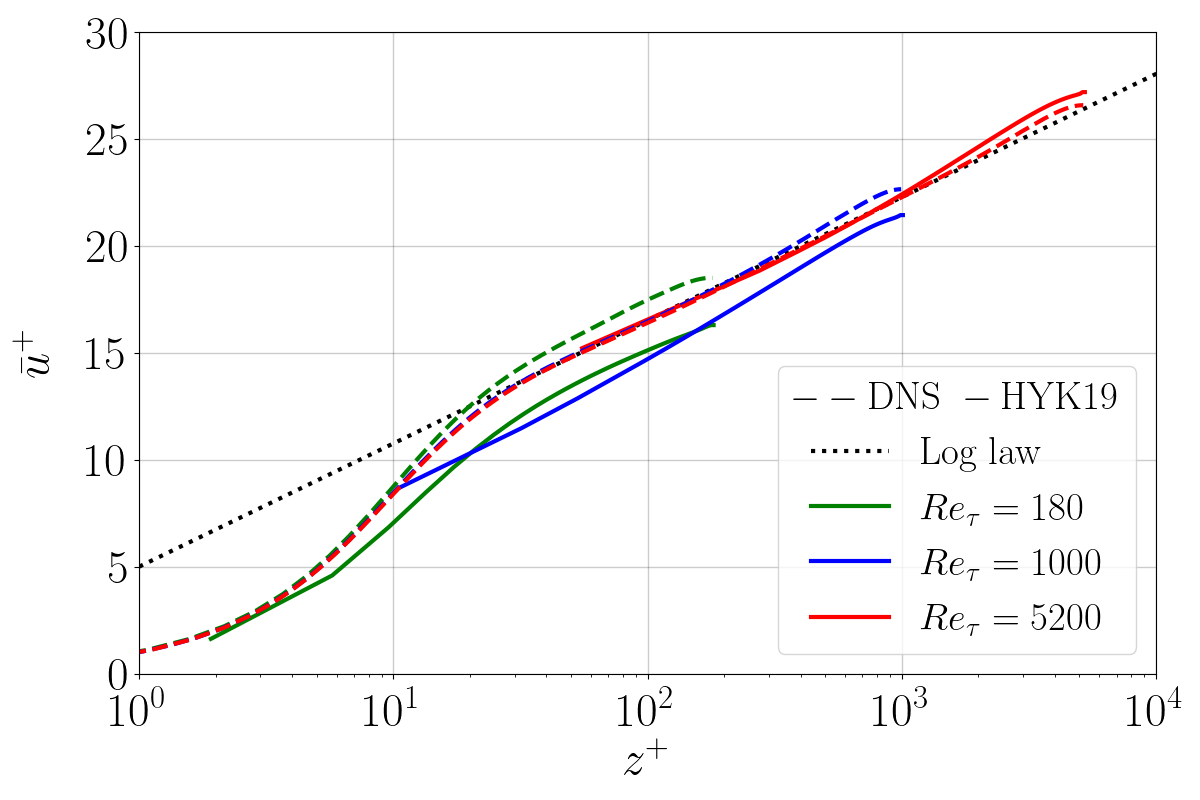}}
\subfigure[ZYZY22.]{\includegraphics[width=0.4\textwidth]{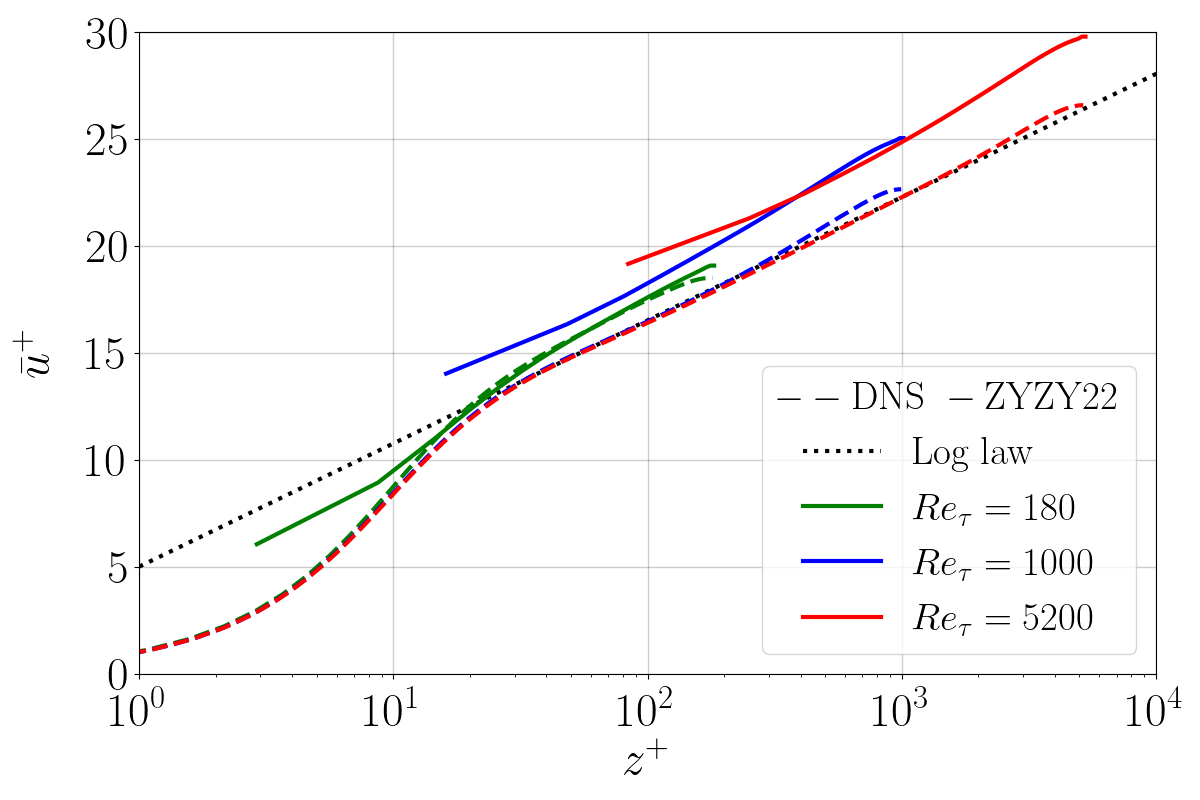}}
\subfigure[BK22.]{\includegraphics[width=0.4\textwidth]{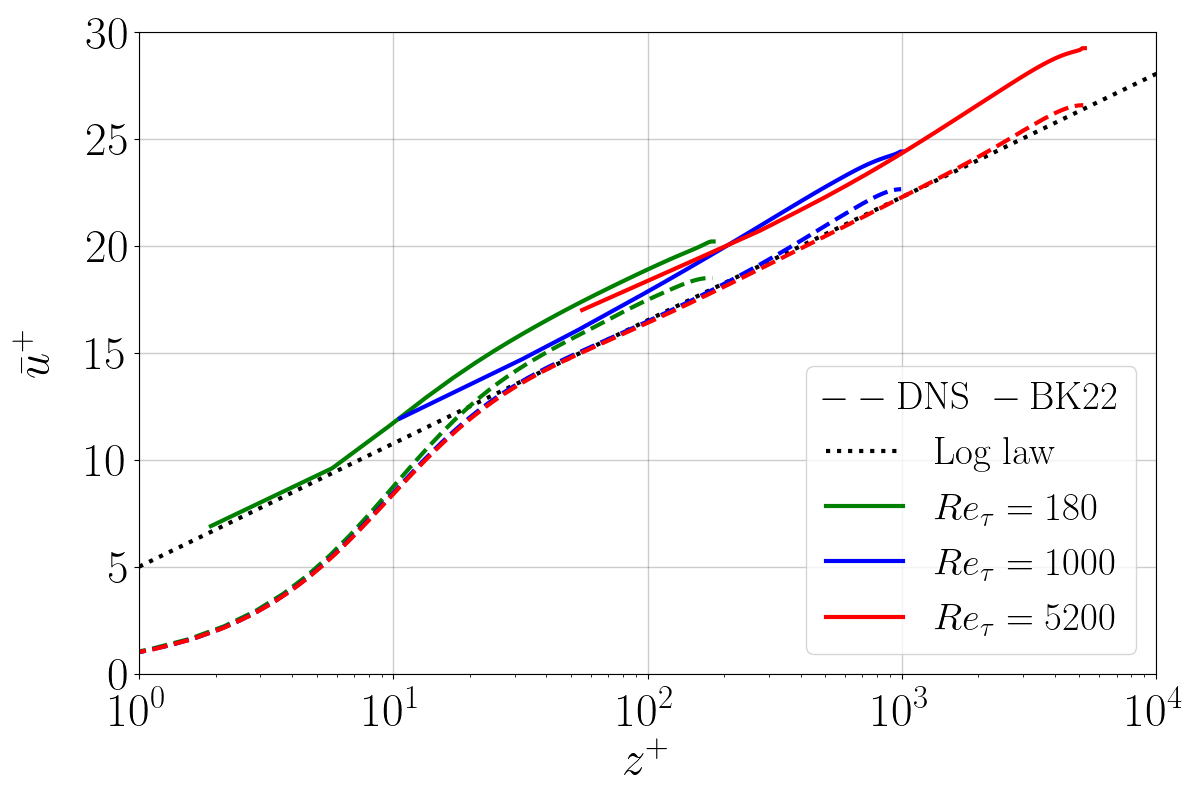}}
\caption{\label{fig:DNS_comp} \black Mean streamwise velocity $\bar{u}^+$ as a function of the wall-normal direction $z^+$ at three Reynolds numbers $Re_\tau=180$, $1000$, and $5200$.
(a) EWM \cite{kawai2012wall}, (b) HYK19 \cite{huang2019wall}, (c) ZYZY22 \cite{zhouArxiv}, (d) BK22 \cite{bae2022scientific}. 
DNS result at $Re_\tau=180$, $Re_\tau=1000$ \cite{graham2016web} and $Re_\tau=5200$ \cite{lee2015direct} are included for comparison purposes.
The log law corresponds to $\kappa=0.4$ and $B=5$.}
\end{figure}

Figure \ref{fig:DNS_comp} compares the mean velocity profiles of DNS and LES for four WMs at three different Reynolds numbers: $Re_\tau=180$, $1000$, and $5200$.
At the largest Reynolds number, EWM and HYK19 WMs perform well as the matching location is located in the log layer, unlike BK22 and ZYZY22 WMs that produce a positive mismatch as seen in Figure \ref{fig:res}.
However, at lower Reynolds numbers, EWM still follows the log law while the matching point is located in the viscous layer at $Re_\tau=180$ and in the buffer layer at $Re_\tau=1000$, leading to a positive mismatch especially at $Re_\tau=180$.
On the other hand, HYK19 WM manages to reproduce the DNS profile reasonably well due to the variation of the intercept according to Figure \ref{fig:interc}, but a negative mismatch remains.

\black

\black
\section{Domain size and SGS model}
\label{app:DS}

\begin{figure}[htbp]
\centering
\subfigure[EWM.]{\includegraphics[width=0.4\textwidth]{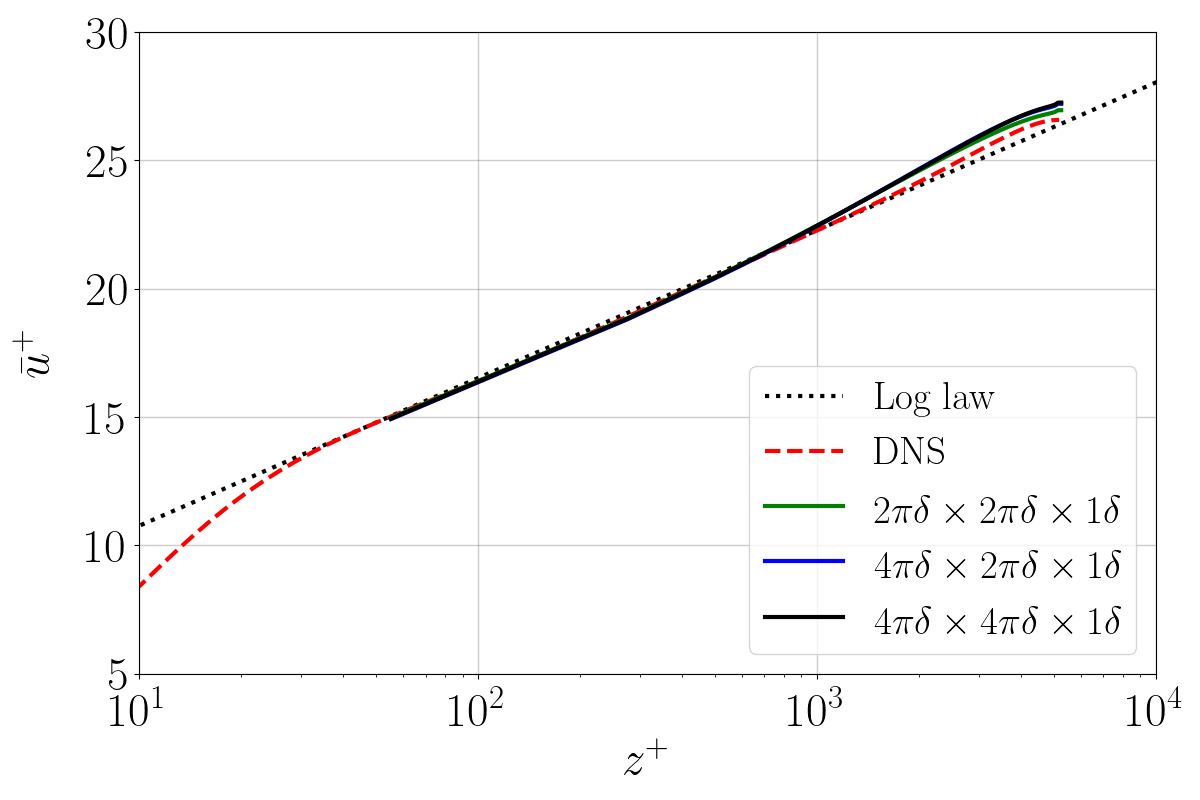}}
\subfigure[HYK19.]{\includegraphics[width=0.4\textwidth]{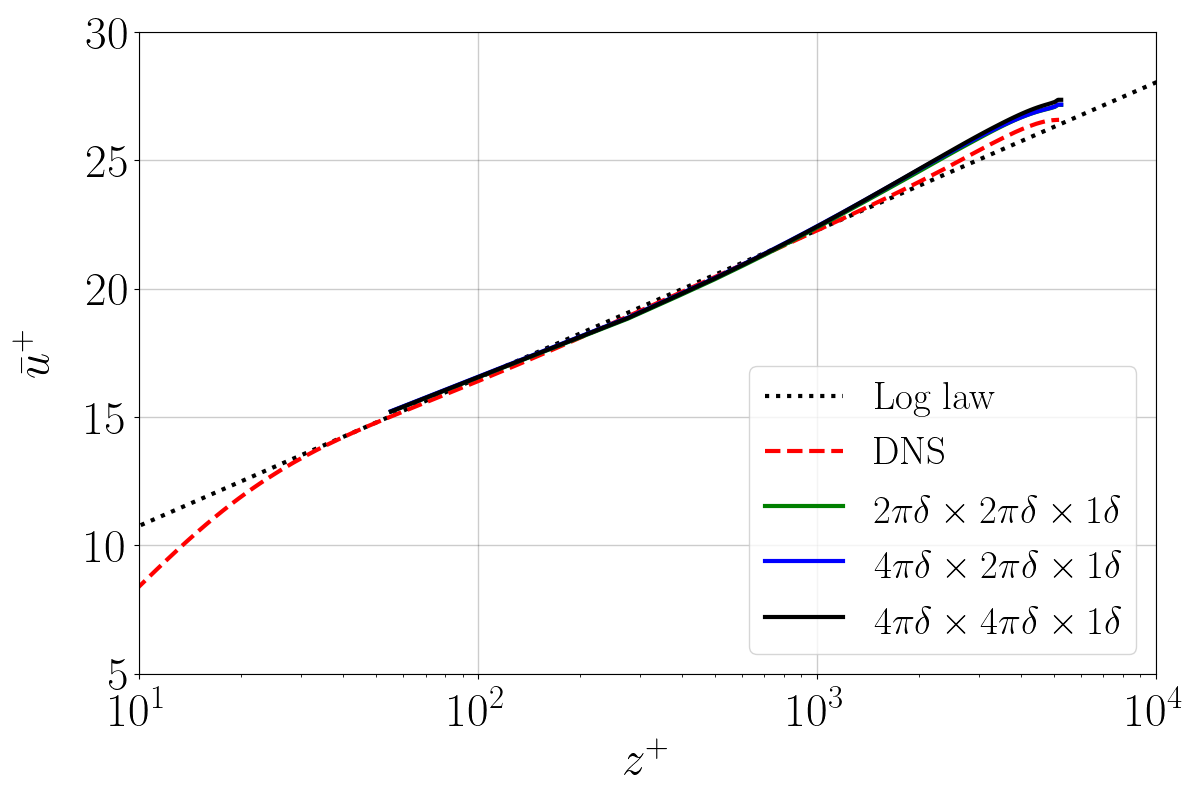}}
\caption{\label{fig:DS}\black Mean velocity $\bar{u}^+$ as a function of the wall-normal coordinate $z^+$ plotted for $Re_\tau = 5200$. 
(a) EWM \cite{kawai2012wall}, (b) HYK19 \cite{huang2019wall}. 
Three different domain sizes are considered keeping the resolution constant: $L_x \times L_y \times L_z = 2\pi \delta \times  2\pi \delta \times 1 \delta$, $4\pi \delta \times 2\pi \delta \times 1 \delta$, and $4\pi \delta \times 4\pi \delta \times 1\delta $.  } 
\end{figure}

In this appendix, it is shown that the effect of computational domain size on the mean velocity for EWM and HYK19 is minimal, as shown in Figure \ref{fig:DS}. 
The baseline computational box size is used along with two additional domain sizes that extend in the spanwise and streamwise directions: $L_x \times L_y \times L_z = 2\pi \delta \times  2\pi \delta \times 1 \delta$, $4\pi \delta \times 2\pi \delta \times 1 \delta$, and $4\pi \delta \times 4\pi \delta \times 1\delta $. 
It is important to note that the resolution is kept constant, so when the domain size is doubled, the number of grid points is also doubled.

\begin{figure}[htbp]
\centering
\includegraphics[width=0.4\textwidth]{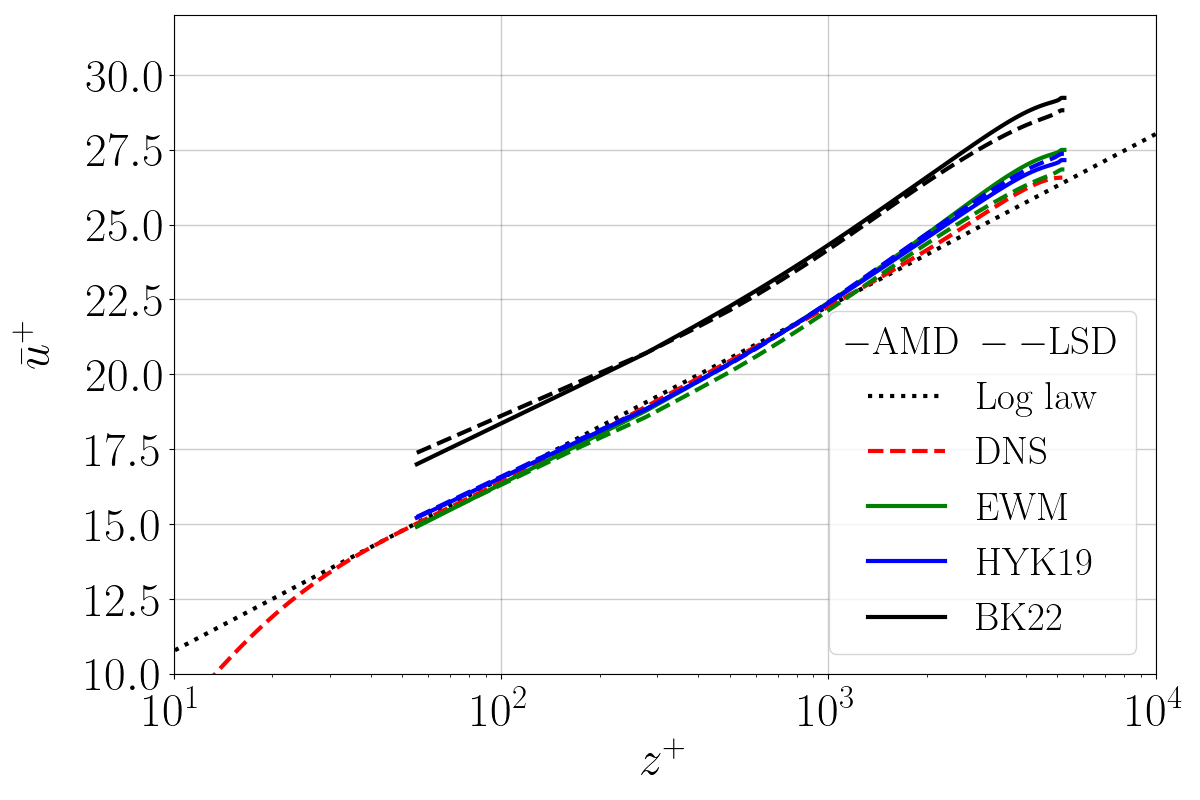}
\caption{\label{fig:SGS}\black Mean velocity $\bar{u}^+$ as a function of the wall-normal coordinate $z^+$ plotted for $Re_\tau = 5200$ using EWM \cite{kawai2012wall}, HYK19 \cite{huang2019wall}, and BK22 \cite{bae2022scientific}. 
Two different SGS models are considered, namely, AMD \cite{rozema2015minimum,abkar2016minimum} and LSD \cite{bou2005scale} models.} 
\end{figure}

We also explored the impact of the SGS model by including the LSD model \cite{bou2005scale}, in addition to the baseline AMD model \cite{rozema2015minimum,abkar2016minimum} in Figure \ref{fig:SGS}.
This topic has been previously investigated in, e.g., Refs. \cite{yang2018hierarchical,wang2020comparative}, among others.
Results show here very little effect of the SGS model for EWM, HYK19 and BK22 at $Re_\tau=5200$.
The effect of the SGS model on MLWMs is similar to that on EWM, indicating that the learned models are not significantly adjusted to compensate for deficiencies in the SGS model.

%\begin{figure}[htbp]
%\centering
%\subfigure[EWM.]{\includegraphics[width=0.4\textwidth]{DS_EWM_rms.png}}
%\subfigure[HYK19.]{\includegraphics[width=0.4\textwidth]{DS_XRWM_rms.png}}
%\caption{\label{fig:DS_rms}\black Root mean square of the streamwise velocity fluctuation %$u_\textrm{rms}^+$ as a function of the wall-normal coordinate $z/\delta$ plotted for $Re_\tau = 5200$.
%(a) EWM \cite{kawai2012wall}, (b) HYK19 \cite{huang2019wall}. 
%Three different domain sizes are considered keeping the resolution constant: $L_x \times %L_y \times L_z = 2\pi \times 2\pi \times 2\pi$, $4\pi \times 2\pi \times 2\pi$, and $4\pi \times 4\pi \times 2\pi$. } 
%\end{figure}

\black

\bibliography{apssamp}% Produces the bibliography via BibTeX.

\end{document}